\documentclass[10pt, draft]{article}
\usepackage{amsmath,amsthm,amsfonts,amscd,eucal,latexsym,amssymb,bm,amsbsy,mathrsfs} 
\usepackage{epsfig}  
\def\sp{\hskip -5pt} 
\def\spa{\hskip -3pt}

\def\cD{{\ca D}}

\def\cN{{\ca N}}
\def\cO{{\ca O}}

\def\cS{{\ca S}}

\def\cW{{\ca W}}
\def\cK{{\ca K}}

\def\sS{{\mathsf S}}

\def\bk{{\bf k}}

\def\mS{\mathscr{S}}  
\def\mF{\mathscr{F}}

\def\bC{{\mathbb C}}           

\def\bN{{\mathbb N}}

\def\bR{{\mathbb R}}

\def\bS{{\mathbb S}}

\newsymbol\rest 1316         

\def\beq{\begin{eqnarray}}
\def\eeq{\end{eqnarray}}
\def\pa{\partial}
\def\at{\left(}               
\def\aq{\left[}               
\def\ag{\left\{}              

\def\ct{\right)}              
\def\cq{\right]}              
\def\cg{\right\}}             
\newcommand{\ca}[1]{{\cal #1}}         

\def\ga{\gamma}
\def\de{\delta}

\def\la{\lambda}

\def\si{\sigma}
\def\om{\omega}

\def\Ga{\Gamma}

\def\Si{\Sigma}

\def\scri{{\Im^{+}}}
\def\scrim{{\Im^{-}}}

\newcommand{\nref}[1]{(\ref{#1})}

\newcounter{proposition}[section]
\newcounter{theorem}[section]
\newcounter{lemma}[section]
\newcounter{definition}[section]
\newcounter{corollary}[section]
\newcounter{remark}[section]
\def\theproposition{\thesection.\arabic{proposition}}
\def\thetheorem{\thesection.\arabic{theorem}}
\def\thelemma{\thesection.\arabic{lemma}}
\def\thedefinition{\thesection.\arabic{definition}}
\def\thecorollary{\thesection.\arabic{corollary}}
\def\theremark{\thesection.\arabic{remark}}

\newcommand{\se}[1]{\section{#1}}

\def\vsp{\vspace{0.2cm}}
\def\vspp{\vspace{0.1cm}}

\def\sse #1 {\vsp\ifhmode{\par}\fi\refstepcounter{subsection}
  \noindent {\bf\thesubsection}. {\em #1}.\quad
  \addcontentsline{toc}{subsection}{\protect\numberline{\thesubsection} #1}%
  }

\def\ssb #1 {\vsp\ifhmode{\par}\fi\refstepcounter{subsection}
  \noindent {\bf\thesubsection.} {\em #1.}\quad
  \addcontentsline{toc}{subsection}{\protect\numberline{\thesubsection} #1}%
  }

\def\ssa #1 {\vsp\ifhmode{\par}\fi\refstepcounter{subsection}
  \noindent {\bf\thesubsection.} {\bf #1.}\quad
  \addcontentsline{toc}{subsection}{\protect\numberline{\thesubsection} #1}%
  }

\def\sssa #1 {\vsp\ifhmode{\par}\fi\refstepcounter{subsubsection}
  \noindent {\bf\thesubsubsection.} {\bf #1.}\quad
  \addcontentsline{toc}{subsubsection}{\protect\numberline{\thesubsubsection} #1}%
  }

\def\proposizione #1 {\vsp\ifhmode{\par}\fi\refstepcounter{proposition}
  \vsp\ifhmode{\par}\fi\noindent {\bf Proposition \theproposition}. \quad {\em #1}}
\def\teorema #1 {\vsp\ifhmode{\par}\fi\refstepcounter{theorem}
  \vsp\ifhmode{\par}\fi\noindent {\bf Theorem \thetheorem}. \quad {\em #1}}
\def\lemma #1 {\vsp\ifhmode{\par}\fi\refstepcounter{lemma}
  \vsp\ifhmode{\par}\fi\noindent {\bf Lemma \thelemma}. \quad {\em #1}}
\def\definizione #1 {\ifhmode{\par}\fi\refstepcounter{definition}
  \vsp\ifhmode{\par}\fi\noindent {\bf Definition \thedefinition}. \quad {\em #1}}
\def\corollario #1 {\vsp\ifhmode{\par}\fi\refstepcounter{corollary}
  \vsp\ifhmode{\par}\fi\noindent {\bf Corollary \thecorollary}. \quad {\em #1}}
  \def\remark {\vsp\ifhmode{\par}\fi\refstepcounter{remark}
  \vsp\ifhmode{\par}\fi\noindent {\bf Remark \theremark}. }

\def\proof #1 {\vspp\ifhmode{\par}\fi\noindent {\it Proof.} {#1} $\Box$\vsp\par}

\begin{document} 
 
\hfill{\sl Desy 08-199, ESI 2088, UTM 726, ZMP-HH/08-21 - December 2008} 
\par 
\bigskip 
\par 
\rm 
 
 
\par 
\bigskip 
\LARGE 
\noindent 
{\bf Distinguished quantum states in a class of cosmological spacetimes and their Hadamard property} 
\bigskip 
\par 
\rm 
\normalsize 
 

\large
\noindent {\bf Claudio Dappiaggi$^{1,a}$},
{\bf Valter Moretti$^{2,b}$}, {\bf Nicola Pinamonti$^{1,c}$} \\
\par
\small
\noindent $^1$ 
II. Institut f\"ur Theoretische Physik, Universit\"at Hamburg,
Luruper Chaussee 149, 
D-22761 Hamburg, Germany.\smallskip\\
\noindent$^2$ Dipartimento di Matematica, Universit\`a di Trento and 
Istituto Nazionale di Alta Matematica -- Unit\`a locale di Trento --
 and  Istituto Nazionale di Fisica Nucleare -- Gruppo Collegato di Trento, \\
via Sommarive 14  
I-38050 Povo (TN), Italy. \smallskip

\noindent E-mail: $^a$claudio.dappiaggi@desy.de,
 $^b$moretti@science.unitn.it,  $^c$nicola.pinamonti@desy.de\\ 
 \normalsize

\par 
 
\rm\normalsize 

\rm\normalsize 
 
 
\par 
\bigskip 

\bigskip
\noindent 
\small 
{\bf Abstract}. 
In a recent paper, we proved that a large class of spacetimes, not necessarily homogeneous or isotropous and
 relevant at a cosmological level, possesses a 
preferred codimension one submanifold, {\it i.e.}, the past cosmological horizon, on which it is possible to 
encode the information of a scalar field theory living in the bulk. Such bulk-to-boundary reconstruction 
procedure entails the identification of a preferred quasifree algebraic state for the bulk theory, enjoying 
remarkable properties concerning invariance under isometries (if any) of the bulk and energy positivity, and reducing 
to well-known vacua in standard situations. 
In this paper, specialising to open FRW models, we extend previously obtained results and we prove that the preferred
state is of Hadamard form, hence the backreaction on the metric is finite and the state can be used as a starting 
point for renormalisation procedures. 
That state could play a distinguished role in the discussion of the evolution of scalar
fluctuations of the metric, an analysis often performed in the development of any model describing the
dynamic of an early Universe which undergoes an inflationary phase of rapid expansion in the past. 
\normalsize
\bigskip 

\noindent Pacs: {\em 04.62.+v, 98.80.Jk}

\se{Introduction}

If one had to carry out a survey in the community of physicists asking for the field of expertise, from which
we can expect in the next few years new exciting and unexpected developments, cosmology 
would be, if not an unanimous, certainly one of the most frequent answers. 

According to the most commonly accepted idea, it is conceivable that, at large scales, a good model of
the geometry of our Universe is given by an homogeneous and isotropic background whose metric is of
Friedmann-Robertson-Walker (FRW) type and whose dynamic is ruled by the Einstein's equation supplemented with
a suitable choice of the ordinary matter. This particular scenario is usually called cold dark matter model. 
Alas, such an approach is not devoid of some flaws and, according to 
modern theoretical cosmology, these can be solved or circumvented assuming that the Universe undergoes an
early phase of rapid expansion, known as the {\em inflation}; moreover such model has the added-on advantage of 
entailing several direct and somehow simple explanations for observed phenomena such as the anisotropies of 
the spectrum of the cosmic microwave background, to quote one, if the most most notable example. A further 
remarkable property of many but not all inflationary models is related to the underlying geometry of the background 
which, at the time of the rapid expansion, turns out to be describable by a de-Sitter spacetime, 
where a very large effective cosmological 
constant appears. 

Although the underlying assumptions of all FRW models, namely homogeneity and isotropy, are 
over-idealisations, it is nonetheless safe to claim that a further distinguished advantage of this line of
reasoning lies in the possibility to explicitly account for inhomogeneities as well as for anisotropies in
terms of suitably modelled perturbations of the metric, they being either of scalar, vector or of tensor form
(there is a vast literature and an interested reader should start looking at \cite{BST, Mukhanov} as well as
at \cite{HWinflation} for a careful discussion of the relations between these mentioned fluctuations and the
inflationary models). Both from a mathematical and from a physical perspective, it is rather interesting to
notice that these perturbations are not supposedly classical but they are truly of quantum nature.
Without entering into  details, they are thought as originating during the phase of rapid expansion as
fluctuations over a suitable ground state and, later, they are let freely evolve. To corroborate such
assumption, it has been shown in the past years how it is possible to retrieve by means of these techniques 
an (almost) scale-free power spectrum, also known as the {\em Harrison-Zeldovich power spectrum}.

Although rather compelling, the above picture cannot be seen as mathematically sound since all the performed
analyses rely on the existence and on the notion of a suitable ground state, a concept which unfortunately is
highly non trivial to set-up for quantum field theories on curved backgrounds. The aim of this paper will be, indeed,
to make this idea precise though we shall only consider free scalar fields thought as a prototype for the 
metric perturbations. Particularly, we will construct and analyse a preferred state in a class of FRW 
spacetimes which are of de Sitter form as the cosmological time $\tau\to -\infty$ in a way we shall better 
specify in the main body of the paper. Most notably we shall show that such a state satisfies the 
{\em Hadamard property} \cite{Wald2,KW}. Roughly speaking and in a physical language it entails both that
the ultraviolet divergences, present at the level of the two-point correlation functions between the fields,
are similar in form to those arising if we deal with the Minkowski vacuum, and that the variance of the 
expectation values of any observable obtained as a limit of coincidence of arguments ({\it i.e.}, Wick 
polynomials of the fields and their derivatives) cannot be divergent \cite{Hollands2}. Therefore such state 
plays the role of a natural starting point in the investigation of interacting field theories by means of 
perturbative techniques. Particularly the stress-energy tensor operator \cite{Mstress}, averaged with 
respect to the state, is finite as well as the backreaction on the metric; therefore the model may be 
gravitationally stable at least at perturbative level (see \cite{DFP} for an example in a cosmological
set up).

The main peculiarity of this paper is that the construction of our proposed state originates from a different 
albeit related scenario which traces its origins back to G. 't Hooft formulation of the holographic principle 
- see \cite{'tHooft} - which advocates the existence of a strong interplay between field theories constructed 
on manifolds of different dimensions. Though often associated either to AdS \cite{Rehren, Duetsch, Ribeiro} 
or to asymptotically flat spacetimes \cite{DMP,M1,DA07}, in \cite{DMP2}, we showed that a bulk-to-boundary 
reconstruction procedure can be successfully set up also in a large class of expanding spacetimes, 
not necessarily homogeneous 
or isotropous, relevant at a 
cosmological level. Assuming homogeneity and isotropy, these encompass, as a subclass,
 the Friedmann-Robertson-Walker spacetimes 
with flat spatial section which satisfy a suitable constraint on the expansion 
factor $a(t)$, namely, as the conformal time $\tau\to -\infty $, the leading behaviour of $a(\tau)$ is that 
of the cosmological de Sitter spacetime.
 
Furthermore, it turns out that all the manifolds satisfying the hypotheses formulated in \cite{DMP2} posses a 
preferred codimension one submanifold, namely, the so-called (past) {\em cosmological horizon} $\Im^-$, on 
which we can encode the information of a scalar field theory on the bulk, barring some further constraints 
both on the mass and on the coupling to scalar curvature. 
As in the asymptotically flat scenario \cite{DMP, 
M1, M2}, we also managed to identify preferred quasifree algebraic state $\lambda$ associated to such 
boundary theory; $\lambda$ satisfies a uniqueness property and, furthermore, it is invariant under a suitable 
notion of asymptotic-symmetry group introduced in \cite{DMP2} for general (not homogeneous or isotropous) expanding universes 
with geodetically-complete past cosmological horizon. $\lambda$ is universal, in the sense that it does not depend 
on the particular bulk spacetime $M$ admitting $\scrim$
as past boundary. However, fixing one of those spacetimes $M$, $\lambda$ can be pulled-back in the bulk $M$ 
to identify a new 
quasifree state $\lambda_M$ which is a natural candidate to play the role of a preferred state associated to 
the bulk theory, since it fulfils several relevant properties: (1) it is invariant under every isometry (if any) of the bulk
which preserves the structure of $\scrim$, (2) $\lambda_M$ admits positive energy with respect to every timelike Killing 
isometry of $M$ (if any) which preserves the structure of $\scrim$, (3) $\lambda_M$ reduces to the well-known Bunch-Davies
vacuum when $M$ is the very de Sitter spacetime.
As already stated, our main goal here is to prove that it is also of Hadamard form.

It is also remarkable to notice that these results could be easily extended to other inflationary scenarios as for example that of the ``power law'' model.
To wit the dynamical content of the underlying quantum fields when considered on the dual Minkowski metric
(see equation \nref{FRWconf1}) 
mimics the one taken here into account.  
In those cases the potential in \nref{rhokeq} which governs the time evolution of the fields with respect to the Minkowski time $\tau$ is still of the form $C\tau^{-2}+O(\tau^{-3})$.
In other words those theories are conformally related to the case investigated in the present paper, hence, since the Hadamard property is invariant under conformal transformations, the definition of a preferred state and the study of its ultraviolet divergences 
could be dealt with in an analogous way.

It is important to notice that there have been previous results tackling the problem to construct 
Hadamard states on FRW spacetimes. It is worth mentioning Olberman's result \cite{Olbermann}, which is based on a previous analysis due to L\"uders and 
Roberts \cite{LR} and to Junker and Schrohe \cite{JS}. 
The difference of our approach is in the choice of a different approximation prescription used in the explicit construction of the 
modes, in the present paper instead of considering the adiabatic
approximation we shall explicitly discuss both the construction of modes out of their asymptotic 
behaviour and the issue of the convergence of the arising perturbative series by means of the so called Green function method.
A further recent result, dwelling in the construction of Hadamard states with nice thermal 
properties is presented in the PhD thesis of K\"usk\"u \cite{Kusku}.

\vskip .2cm

The outline of the paper is the following: in the next section we briefly recollect some of the results of 
\cite{DMP2} and, particularly, we discuss the main geometric properties of the backgrounds we shall consider 
identifying a preferred codimension 1 submanifold, namely, the cosmological horizon. In section 2 we set up 
the bulk-to-boundary reconstruction procedure for a scalar field and we identify a preferred algebraic 
quasifree bulk state, extending some results previously achieved in \cite{DMP2}.
Furthermore, we shall discuss the regularity properties of the 
solutions of the bulk equations of motion once they are restricted on the horizon and once suitable
constraints on the mass $m$ and on the coupling to scalar curvature $\xi$ are imposed. 
In the third section we prove the main result of the paper, {\it i.e.}, the two-point function associated to 
the constructed state in the bulk is a well-defined distribution of Hadamard form. Eventually in the fourth section,
we briefly draw some conclusions.

\ssa{Expanding Friedmann-Robertson-Walker spacetime with flat spatial sections}
Let us remind the reader some
 geometric properties of the spacetimes we intend to consider. As this topic has been
dealt with in greater details in \cite{DMP2}, in this section we shall present only 
the ingredients relevant to understand the main statements of this paper,  while pointing a reader interested 
in further discussions and details to \cite{DMP2}.\\
The homogeneous and isotropic solutions of Einstein's equation which are of cosmological interest
can be described as a four dimensional smooth Lorentzian manifold $M$ equipped with the following metric:
\beq\label{FRW}
g=-dt^2+a^2(t)\left[\frac{dr^2}{1-kr^2}+r^2d\bS^2(\theta,\varphi)\right].
\eeq
Here $k$ takes the values $-1,0,1$ and it indicates whether the constant time 
hypersurfaces are respectively hyperbolic, flat or parabolic, whereas $a(t)$ is a smooth function of
constant sign depending only on the variable $t$, whose domain ranges in an open interval $I=(\alpha,\beta)$.
Such class of backgrounds is too large for our purposes and, therefore, we shall henceforth restrict our 
attention to the subclass with both $k=0$ and $\dot{a}(t)\geq 0$. It represents, at a physical level, a
remarkably interesting case of a globally-hyperbolic Friedmann-Robertson-Walker (FRW) expanding Universe 
$(M,g_{FRW})$ with flat spatial sections. $M$ is diffeomorphic to $I \times \bR^3$ and, up to a change of 
coordinates, the metric reads
\beq\label{FRWconf1}
g_{FRW}\doteq a^2(\tau)\left[-d\tau^2+{dr^2}+r^2d\bS^2(\theta,\varphi)\right].
\eeq
Here $\tau$ is the so-called {\bf conformal time} constructed out of the defining identity $a(\tau) d\tau
\doteq dt$.
As a last restriction, we require the conformal factor to have the following form referring to its behaviour 
as $\tau \to -\infty$:
\beq\label{const}
a(\tau)=  -\frac{1}{H\; \tau} +O\at\frac{1}{\tau^2}\ct \;, \quad
\frac{d a(\tau)}{d\tau}=\frac{1}{H\; \tau^2} +O\at\frac{1}{\tau^3}\ct,\quad\frac{d^2 a(\tau)}{d\tau^2}=-\frac
{1}{H\;\tau^3} +O\at\frac{1}{\tau^4}\ct.
\eeq
The reason underlying such constraints is twofold. On the one hand they identify a class of physically
relevant spacetimes with a rather distinguished geometric property, namely they all posses a preferred
codimension one submanifold which represents the natural screen on which to encode the data of a bulk field
theory. On the other hand \eqref{FRWconf1} and \eqref{const} characterise those backgrounds which look
``asymptotically'' - {\it i.e.}, as $\tau\to - \infty$ - as the cosmological de Sitter Universe, whose 
expansion factor in \eqref{FRWconf1} is 
$$a_{dS}(\tau)\doteq -\frac{1}{H\; \tau},$$
and  $\tau \in I =(-\infty,\beta)$. This property is shared also by
all others solutions of \eqref{FRWconf1} satisfying \eqref{const}. Above $H$ is the so-called 
{\bf Hubble parameter} and we felt safe to adopt the same symbol also in the preceding formulas.\\
The class of spacetimes we characterised has a quite remarkable application in the description of the early 
stages of the evolution of the universe, since most of the so-called inflationary scenarios are based on a 
phase of rapid expansion modelled by a scalar field on an (asymptotically) de Sitter background. 

\ssa{Past cosmological horizons}
As  we discussed in detail in \cite{DMP2}, all the
spacetimes under analysis  are globally hyperbolic -- the constant-$\tau$ hypersurfaces being trivially 
smooth spacelike Cauchy surfaces -- and they posses a boundary made of a null codimension one-submanifold 
$\scrim$ on which to encode the data of a bulk field-theory. This is the so-called ({\bf past}) {\bf 
cosmological horizon} as defined by Rindler in \cite{Rindler}. To concretely characterise $\scrim$, first one 
performs the coordinate change $U=\tan^{-1}(\tau-r)$, $V=\tan^{-1}(\tau+r)$; in this way one can realize by 
direct inspection that $(M,g_{FRW})$ could be read as an open submanifold of a larger spacetime 
$(\widehat M,\widehat g)$, {\it i.e.}, $M\subset\widehat M$ and $\widehat g|_M\equiv g_{FRW}$. In this
framework,  $\Im^-$ is nothing but the past causal boundary of $M$ in $\widehat{M}$, that is 
$M = I^+(\scrim; \widehat{M})$ and
$\Im^-=\partial M=\partial J^{+}
(M,\widehat M)$.  $\Im^-$ turns out to be diffeomorphic to $\bR\times\bS^2$ as well as a null differentiable
manifold. We shall also exploit some further remarkable geometric properties valid in a neighbourhood of 
$\scrim$, namely $a(\tau(U,V))$ vanishes identically on $\Im^-$ whereas the differential $da(U,V)|_{\Im^-}=-2
H^{-1} dV$ does not. Such feature entails that the metric $\widehat g$ can be restricted on the horizon where
it picks up a (geodetically complete) {\bf Bondi-like form}:
\beq\label{Bondiform}
\widehat{g}|_{\Im^-}= H^{-2} \left(-2d\ell da+d\bS^2(\theta,\varphi)\right)\:.
\eeq
Here $d\bS^2$ is the standard metric on the $2$-sphere, whereas, up to a constant, $\ell(V)=-H\tan(V)$,
is the affine parameter of the integral line of the vector $n\doteq\nabla a$ which turns out to be a complete 
null geodesics of $\widehat{g}$. Furthermore the above
remark on the form of the expansion factor and of its derivative on the horizon also leads to conclude 
$\mathcal{L}_{\partial_\tau}\widehat g= -2\partial_\tau\left(\ln a\right)\widehat g,$
where the right hand side vanishes on $\Im^-$ as $a$ does. Hence we can infer that $\partial_\tau$ is a 
conformal Killing vector which, approaching $\Im^-$, both tends to become tangent to it and to coincide with 
$-H^{-1} \widehat\nabla^b a$. Such property will come in very handy in the forthcoming discussions.

\remark
\label{nic} In this paper we shall confine ourselves to FRW models. 
However, we stress that, as discussed in \cite{DMP2}, most results presented therein  are valid also for 
spacetimes which are neither homogeneous nor isotropous, when they admit a geodetically-complete past 
cosmological horizon and a preferred conformal time $\partial_\tau$ with the previously discussed interplay 
with the geometry of the horizon. From a very abstract point of view, as established in \cite{DMP2}, $\scrim$
can be equipped with a certain {\em infinite-dimensional} group of isometries $SG_{\Im^-}$ which is the 
analogue of the BMS group for asymptotically flat spacetimes \cite{DMP}. This group, on the one hand depends on 
the only structure of $\scrim$, hence in this sense is universal, while, on the other hand, it embodies -- 
through a faithful representation --  all possible bulk Killing symmetries which preserve the structure of 
$\scrim$ of every -- {\em not necessarily homogeneous or isotropous} -- cosmological model which admits 
$\scrim$ as a boundary (see proposition 3.2 as discussed in \cite{DMP2}).\\
Also in the general case considered in \cite{DMP2}, $\scrim$ is a natural candidate on which to encode the 
information of bulk data of the scalar free QFT. As we shall shortly discuss in the particular case 
considered in this paper, it has been proved in \cite{DMP2} that it is possible both to construct a genuine 
free scalar quantum field theory defined on $\Im^-$ and to associate to it a preferred state enjoying 
invariance under $SG_{\Im^-}$. Such theory can be induced back to the bulk giving rise to a second one which 
turns out to be automatically invariant under every symmetry of the bulk which preserves $\scrim$.

\se{From the bulk to the boundary and back} 
The aim of this subsection is to sketch the scheme of quantisation of a free scalar field living on any of 
the spacetimes under analysis, as previously discussed in \cite{DMP2}, focusing on how it is possible to 
associate every quantum theory living in $(M,g_{FRW})$ with a dual theory on $\Im^-$. The main advantage of 
pursuing such approach lies in the existence of a preferred quasifree state (see remark \ref{nic}) for the 
boundary theory that we shall eventually pullback to the bulk, picking out a natural preferred quasifree 
state $\lambda_M$ for the QFT in the spacetime $(M,g_{RFW})$. The state $\lambda_M$ generalises the Bunch-Davies
vacuum for more general expanding universes. The analysis of the remarkable properties of $\omega_M$ was 
started in \cite{DMP2} for a wider class of spacetimes (dropping the requirements of homogeneity and 
isotropy); here we only focus on the validity of the Hadamard property for $\lambda_M$ in the class of RFW 
spacetimes we are considering. 

Since  $(M,g_{FRW})$ is globally hyperbolic, the Cauchy problem for smooth compactly-supported initial data is 
well-posed
\cite{Wald2,BGP}.
Let us thus consider the Klein-Gordon equation for the real scalar field $\Phi$ arbitrarily coupled $\xi$ 
with the scalar curvature:
\beq\label{wave}
P\Phi=0,\quad\textrm{where}\quad P=-\square+\xi{R}+m^2\:,
\eeq
where $\square$ is the D'Alembert operator associated with $g_{FRW}$ and $m^2\geq 0$. The space of real 
smooth solutions $\mathcal{S}(M)$ of (\ref{wave}) with compactly supported smooth Cauchy data is a 
symplectic space $(\mathcal{S}(M),\sigma_M)$ when endowed with the Cauchy-surface independent nondegenerate 
symplectic form:
\beq\label{sympla}
\sigma_M(\Phi_1,\Phi_2) \doteq \int\limits_\Sigma
d\mu_{g_{FRW}}^{(\Sigma)}\left(\Phi_2\nabla_n\Phi_1-\Phi_1\nabla_n\Phi_2\right),\quad\forall\Phi_1,\Phi_2\in
\mathcal{S}(M)\:.
\eeq
Above, $\Sigma$ is an arbitrary spacelike smooth Cauchy surface, $d\mu_{g_{FRW}}^{(\Sigma)}$ the  measure on
$\Sigma$ induced by the metric and $n$ is the unit future-pointing vector orthogonal to $\Sigma$. 
It is then a standard procedure \cite{KW,Wald2,BR2} to associate $(\mathcal{S}(M),\sigma_M)$ with the Weyl
$C^*$-algebra $\cW(M)$, determined up to isometric $*$-isomorphisms and constructed out of the Weyl 
generators $W_M(\Phi)\neq 0$, satisfying, for all $\Phi\in\mathcal{S}(M)$, the {\bf Weyl relations}
\beq\label{Weyldef1}
W_M(\Phi) = W_M^*(-\Phi),\qquad W_M(\Phi) W_M(\Phi')=e^{\frac{i}{2}\sigma_M(\Phi,\Phi')}
W_M(\Phi+\Phi').
\eeq
The self-adjoint elements of $\cW(M)$ represent the quantum observable of the bosonic free quantum field 
theory of the field $\Phi$ and hence $\cW(M)$ realizes the quantisation of the theory at algebraic level 
\cite{Wald2, KW, BR, BR2}.

\ssa{Modes} \label{secmodes} We wish now to better characterise $\mathcal{S}(M)$ employing constant-time hypersurfaces 
$\Sigma_\tau$ (which are diffeomorphic to $\bR^3$) as Cauchy surfaces and adopting standard Cartesian 
coordinates $(\tau,\vec{x})$ on $M$ as in \nref{FRWconf1}.
Adopting the convention that $\bk\in\bR^3$ and $k=|\bk|$, a generic element $\Phi\in\mathcal{S}(M)$ can be 
decomposed as:
\beq\label{general}
\Phi(\tau,\vec{x})=\int\limits_{\bR^3}d^3\bk\left[\phi_{\bf k}(\tau,\vec{x})\widetilde\Phi(\bk)+\overline{
\phi_{\bf k}(\tau,\vec{x})\widetilde\Phi(\bk)}\; \right]\quad \mbox{with}\quad
\phi_{k}(\tau,\vec{x})\doteq \frac{e^{i{\bf k}\cdot \vec{x}}}{(2\pi)^{\frac{3}{2}}}\frac{\chi_{k}(\tau)}
{a(\tau)}\;.
\eeq
The modes $\phi_{\bf k}$ are constructed out of the $\chi_k(\tau)$, which are solutions of the differential 
equation:
\begin{gather}
 \frac{d^2}{d\tau^2}\chi_k(\tau)  + (V_0(k,\tau)  + V(\tau)) \chi_k(\tau) =0,  \nonumber \\ 
 \quad
V(\tau)\doteq 
 k^2 + a(\tau)^2\left[m^2 + \left(\xi- \frac{1}{6} \right)R(\tau) \right] - V_0(k,\tau)\:, \label{rhokeq}
\end{gather}
where $V_0(k,\tau) \doteq  
k^2 \spa +a_{dS}(\tau)^2\spa\left[m^2 + \left(\xi- \frac{1}{6} \right)12 H^2\right]$,
so that it results $V(\tau)= O(1/\tau^3)$ in view of (\ref{const}) as $\tau \to -\infty$. 
The following  condition is also assumed:
\beq\label{constraint}
\frac{d\overline{\chi_{k}(\tau)}}{d\tau}\chi_{k}(\tau)-\overline{\chi_{k}(\tau)}\frac{d\chi_{k}(\tau)}{d\tau}
=i\:, \quad \tau\in \bR_-\:.
\eeq
We now define: 
\beq\label{nu}
\nu =\sqrt{\frac{9}{4}- \at \frac{m^2}{H^2}  + 12 \xi \ct}\;, \quad \mbox{\em where we always assume  both $Re \nu \geq 0$ and   $Im \nu \geq 0$,}
\eeq
noticing that $\nu$ can be either real or imaginary, but not a general complex number. 
A general solution of \eqref{rhokeq} satisfying the constraint (\ref{constraint}) and with $Re\nu < 1/2$
can be constructed as a convergent series, as discussed in 
Theorem 4.5 of \cite{DMP2}. Therein,  $V(\tau)$ is treated as a perturbation potential over the solutions 
\eqref{mode} in de Sitter background where $V\equiv 0$. 
For the purely de Sitter spacetime ($V\equiv 0$) solutions of \eqref{rhokeq} satisfying the constraint (\ref{constraint})
are fixed to be:
\beq\label{mode}
\chi_{k}(\tau)=\frac{\sqrt{-\pi\tau}}{2}e^{\frac{-i\pi\nu}{2}}\overline{H^{(2)}}_\nu(-k\tau),
\eeq
where $H^{(2)}_\nu$ is the Hankel function of second kind \cite{SS}. Moreover the perturbative procedure to 
construct the modes $\chi_k(\tau)$ for the general background yields the constraints (valid for the pure de 
Sitter case, too):
\beq \label{infinity} \lim_{\tau\to -\infty}\sp \chi_k(\tau)e^{i k \tau } 
= \frac{e^{-i\pi/4}}{\sqrt{2 k}}\,,
\quad 
\lim_{\tau\to -\infty} \sp\frac{d \chi_k}{d\tau}
(\tau)
e^{i k \tau}  = - i e^{-i\pi/4}\sqrt{\frac{k}{2}}\:.  \eeq
\remark\label{remarkk} 

{\bf (1)} The perturbative construction of the smooth real solutions of the  Klein-Gordon equation
with compactly supported data, as presented in Theorem 4.5 in \cite{DMP2}, extends with minor changes
to the case 
$Re\nu < 3/2$ provided that the potential decays as $V(\tau)= O\at\frac{1}{\tau^5}\ct$, which corresponds 
to the stricter constraints on the rate of expansion $a(\tau)$:
\beq\label{const2}
a(\tau)=  -\frac{1}{H\; \tau} +O\at\frac{1}{\tau^{5/2}}\ct \;, \quad
\frac{d a(\tau)}{d\tau}=\frac{1}{H\; \tau^2} +O\at\frac{1}{\tau^3}\ct,\quad\frac{d^2 a(\tau)}{d\tau^2}=-\frac
{1}{H\;\tau^3} +O\at\frac{1}{\tau^6}\ct.
\eeq
Nonetheless we feel that, despite such further restriction on the choice of the underlying geometry,
it is interesting from a physical perspective to allow  $\nu$ to be as close to $\frac{3}{2}$ as possible. As
a matter of facts, in this case the power spectrum $P(k,\tau)$ of the scalar field will be close to the 
{\em Harrison-Zeldovich scale} free one. As an evidence of this claim, notice that, at least at small scales and as
$\tau \to -\infty$,
$$
P(k,\tau)= \overline{\chi_k(\tau)}\chi_k(\tau)\sim \frac{1}{|k|^{2\nu}}
$$
as one can infer from the analysis performed in the appendix \ref{approxsol}. If we set $\nu =3/2$, then we
end up with a genuine scale free spectrum.

{\bf (2)}  The identity \eqref{general} inverts as:
\beq\label{modes}
\widetilde\Phi(\bk)=-i\int\limits_{\Si_{\tau}}d^3x\; a^2(\tau)\;\left[
\frac{\partial\overline{\phi_{\bf k}(\tau,\vec{x
})}}{\partial\tau}\Phi(\tau,\vec{x})-
\overline{\phi_{\bf k}(\tau,\vec{x})}\frac{\pa\Phi(\tau,\vec{x})}{\partial
\tau}\right],
\eeq
$\Sigma_{\tau}$ being any constant-time hypersurface. As the right hand side is independent
from the choice of a specific value $\tau$, we are free to let $\Sigma_{\tau}$ coincide with the Cauchy 
surface. In this way, $\Phi(\tau,\vec{x})$ and $\frac{\pa \Phi}{\pa\tau} (\tau,\vec{x
})$ are the assigned initial data of the considered element $\Phi \in \cS(M)$ individuated by $\widetilde\Phi$
inserted in the right-hand side of (\ref{general}).

 {\bf (3)} Out of the behaviour of Hankel functions in a neighbourhood of the 
origin and in the perturbative construction of the general solution valid both for $Re\nu < 1/2$ and 
$V(\tau)= O\at\frac{1}{\tau^3}\ct$, or for $Re\nu < 3/2$ and 
$V(\tau)= O\at\frac{1}{\tau^5}\ct$ (see in the appendix \ref{approxsol} for more details), the shape of $\widetilde{\Phi}$ as $\bk\to 0$ is as follows: whether 
$\nu$ is imaginary, no singularity occurs, whereas, if $Re \nu > 0$, following the analysis 
performed in the proof of Theorem 4.5 in \cite{DMP2} (and its extension to the case
$Re \nu < 3/2$, in appendix \nref{approxsol}), one gets that
\beq\label{uniform}
|\partial^n_k\widetilde{\Phi}({\bf k})| \leq C_{\delta,n}/|{\bf k}|^{Re\nu+n}\:,\quad \mbox{for $0<|\bk| 
\leq \delta$,}
\eeq
if $n=0,1,2,3$ and for some $C_{\delta,n}, 
\delta >0$. These estimates arise for the analogous of the functions $k\mapsto\chi_k(\tau)$, $k\mapsto
\partial_\tau\chi_k(\tau)$ and their $k$-derivatives.\\
For $|\bk|\to +
\infty$ we have instead the following behaviour.
As the $k \mapsto \chi_k(\tau)$, $k \mapsto \partial_\tau\chi_k(\tau)$ and their $k$-derivatives increase at 
most polynomially and since the Cauchy data are smooth and compactly supported, (\ref{modes}) entails 
that $\widetilde\Phi \in C^\infty(\bR^3 \setminus \{0\}; \bC)$ and,
for every $\Delta>0$, $n,m=0,1,2,...$ there are constants $B_{\Delta,n,m}$ with:
\beq\label{uniform1}
|\partial^n_k\widetilde{\Phi}({\bf k})| \leq B_{\Delta,n,m}/|{\bf k}|^{m}\:,\quad \mbox{for $|\bk| 
\geq \Delta$.}
\eeq

\ssa{Projection of the quantum theory on the Horizon}\label{secproj}
Let us now focus our attention on the horizon itself. Since $\Im^-$ is diffeomorphic to $\mathbb{R}
\times\bS^2$, we adopt the coordinates $(\ell,\theta,\varphi) \in \bR \times \bS^2$ used in (\ref{Bondiform}).  
We want to define a suitable symplectic space in order to construct the Weyl algebra of the observables 
defined on a null surface as $\scrim$ \cite{MP4,DMP,DMP2}. To this end, the introduction of some useful 
mathematical tools is in due course.\\
The complex smooth functions which decay, with every derivative, faster than every negative power of $\ell$ 
uniformly in the angular variables) will be indicated by $\mS(\bR\times\bS^2)$. Notice that, if $f \in \mS(\bR\times\bS^2)$, then
$f(\cdot, \omega) \in \mS(\bR)$ for every fixed $\omega \in \bS^2$.
The complete dual space (with respect to the natural Frech\'et topology) of $\mS(\bR\times\bS^2)$ will be 
denoted by  $\mS'(\bR\times\bS^2)$. In the following $\widehat{\psi}$ denotes the {\bf Fourier transform}
\footnote{All that follows is a very straightforward extension of the standard theory of Fourier transform.
Further details were presented in the Appendix C of \cite{M2} where we used complex coordinates $(\zeta,\bar{
\zeta})$ on the sphere instead of our $(\theta,\varphi)$, but this affects by no means the definitions and 
results.} of the distribution $\psi\in \mS'(\bR\times\bS^2)$. As in the standard theory, this transformation
is defined by assuming that, if $\phi \in \mS(\bR\times\bS^2)$, 
\beq
\widehat{\phi}(k, \omega) \doteq \int_{\bR} \frac{e^{ik\ell}}{\sqrt{2\pi}} {\psi}(\ell, \omega) d\ell\:, 
\quad \forall (k,\omega) \in \bR \times \bS^2\:, \label{fourier}
\eeq
so that $\widehat{\phi} \in \mS(\bR\times\bS^2)$ again,
and afterwards, extending the definition per duality to $T \in \mS'(\bR\times\bS^2)$ as
$\langle\: \widehat{T}, \phi \rangle \doteq \langle T, \widehat{\phi} \:\rangle$ for all $\phi \in \mS(\bR
\times\bS^2)$.
The Fourier transform turns out to be bijective and continuous both
as a map $\mS(\bR\times\bS^2)\to \mS(\bR\times\bS^2)$ and $\mS'(\bR\times\bS^2)\to \mS'(\bR\times\bS^2)$
whereas the inverse transform is obtained by duality starting from the {\bf inverse Fourier transform} on 
$\mS(\bR\times\bS^2)$:
\beq
 {\psi}(\ell, \omega)\doteq \int_{\bR} \frac{e^{-ik\ell}}{\sqrt{2\pi}} \widehat{\phi}(k, \omega) d\ell\:, \quad \forall (\ell,\omega)
 \in \bR \times \bS^2\:, \label{fourier2}
\eeq
If $\phi \in L^1(\bR\times\bS^2,d\ell dS^2)$, so that
$\phi \in \mS'(\bR\times\bS^2)$, its Fourier transform, can be equivalently computed 
as the right-hand side of (\ref{fourier}) and $\widehat{\phi}$ is $k$-continuous.\\
Using these tools,
as a first step to define a bosonic field theory, we  introduce the symplectic 
space of real wavefunctions $(\cS(\Im^-),\sigma)$ relaxing the requirements on the elements of the space with respect to
that done in  \cite{DMP2} in order to encompass the physically interesting case $Re \nu < 3/2$ as we shall see shortly:
\begin{gather} 
\cS(\Im^-)\doteq \left\{\left.\psi\in C^\infty(\bR\times\bS^2)\;\right|\; ||\psi||_\infty\:, ||k \widehat{
\psi}||_\infty < \infty\:, \partial_\ell\psi\in
L^1(\bR\times\bS^2,d\ell dS^2)\:,
\widehat{\psi} \in
L^1(\bR\times\bS^2,dk dS^2)
\right\}, \label{SS}\\
\sigma_{\scrim}(\psi,\psi')\doteq \int\limits_{\bR\times\bS^2}\left(\psi\frac{\partial\psi'}{\partial \ell}-
\psi'\frac{\partial\psi}{\partial \ell}\right) d\ell dS^2\:, \quad\forall\psi,\psi'\in S(\Im^-)\:.
\label{sympl}
\end{gather}
Notice that above, where $\psi\in C^\infty(\bR\times\bS^2)$ and it is bounded, the Fourier transform $\widehat{\psi}$ makes 
sense in the distributional sense. In \cite{DMP2},  $\cS(\scrim)$ was defined as the space of smooth real-valued functions 
of $L^2(\bR \times \bS^2; d\ell dS^2)$ with $\ell$-derivative in $L^2(\bR \bS^2; d\ell dS^2)$.
In that case the Fourier transform could be interpreted  as a Fourier-Plancherel transform.
In our case this is not possible in general. \\ 

\noindent Since $\sigma_{\scrim}$ is non-degenerate, it is possible to associate to $(S(\Im^-),\sigma_{\scrim})$ a unique, up to 
isometric  $*$-isomorphism, Weyl $C^*$-algebra $\mathcal{W}(\Im^-)$ whose generators 
$W_{\Im^-}(\psi)\neq 0$ for $\psi\in S(\Im^-)$ satisfy the {\it Weyl  relations} (\ref{Weyldef1}) with 
$W_M$ replaced by $W_{\scrim}$,
$\sigma_M$ replaced by $\sigma$ and $\Phi,\Phi'$ replaced by $\psi,\psi'$.
As for the bulk, $\mathcal{W}(\Im^-)$ represents a well-defined set of basic observables and, hence, it can 
be thought as the building block of a full-fledged quantum scalar field theory on the cosmological horizon 
$\scrim$. \\
Nonetheless such line of reasoning would be spurious if we were not able to connect the information
arising from the boundary to the bulk counterpart. In \cite{DMP2}, we 
 tackled this problem showing that it is possible to realize
 $(\cS(M),\si_M)$ as a subspace of $(\cS(\scrim),\si_{\scrim})$  by means of
an injective symplectomorphism $\cS(M) \to \cS(\scrim)$. This result, in turn, 
implies the existence 
of an identification, $\imath : \cW(M) \to \cW(\scrim)$, of the algebra of bulk observables $\mathcal{W}(M)$ 
and a sub algebra 
of observables of the boundary $\mathcal{W}(\Im^-)$. We review the procedure showing, in  theorem \ref{projb} below,
 that the result 
is valid also with our more general definition of  $\cS(\Im^-)$ and referring to a large class of values of $\nu$
which includes the most physically interesting ones as stressed in (1) of remark \ref{remarkk}.

We start by reminding that any of the spacetimes $M$ we are considering can be extended to a second 
spacetime $\widehat M$ which both is globally hyperbolic in its own right and it includes $\Im^-$ as a null 
hypersurface \cite{DMP2}. By a standard argument (see \cite{BS06} for the general case), outside the support of Cauchy data of $\Phi$, one can deform the 
employed Cauchy surface of $M$ to a Cauchy surface of $\widehat M$. 
Since $P$ is a second-order hyperbolic partial differential operator and it can be extended in the analogy for
$\widehat M$, a unique solution $\Phi'$ of  \eqref{wave} exists in $\widehat M$ with the same initial 
compactly supported data  as those of $\Phi$. By uniqueness $\Phi'\spa\rest_M =\Phi$. Furthermore, since also
$\Im^-\subset \widehat M$, we can define the linear map:
\beq\label{proje}
-H^{-1} \Gamma:\mathcal{S}(M)\to C^\infty(\Im^-; \bR) \quad \mbox{such that $\Gamma(\Phi)\doteq\Phi'\spa 
\rest_{\Im^-}$.}
\eeq
However such a result does not guarantee a sufficient
regularity of the image on $\Im^-$ of the solution of \eqref{wave} in order that $\Gamma(\Phi)\in\cS(
\scrim)$; therefore, we shall analyse more in detail the structure of $\Phi$ itself. The following technical 
proposition establishes in fact that this is the case. It is based on the following observation.
With the same procedure preformed in the proof of theorem 4.4 in \cite{DMP2}, employing the estimate stated in (3)
of remark \ref{remarkk},
 one gets that:
\beq\label{proj}
-\frac{1}{H}\left(\Gamma\Phi\right)(\ell,\theta,\varphi)=iHe^{-i\frac{\pi}{4}}\int\limits_0^\infty\frac{e^{-i
\ell k}}{\sqrt{2\pi}}\sqrt{\frac{kH}{2}}\widetilde\Phi(Hk,\eta(\theta,\varphi)) dk + c.c., 
\eeq
where $\widetilde\Phi$ coincides with \eqref{modes} written in suitable spherical coordinates ({\it i.e.,}
$\widetilde{\Phi}({\bf u})= \widetilde{\Phi}(|{\bf u}|, \vartheta, \phi)$, with $(\vartheta, \phi)$ polar 
angles of ${\bf u}$) and 
$\eta: (\theta,\varphi) \mapsto (\pi -\theta, \varphi+ 2\pi)$ the parity inversion.

\proposizione{\label{decay}  
Assume that $\xi$ and $m$ are such that either $\nu$ in \eqref{nu} satisfies $ Re \nu<3/2$, and $V(\tau)=O(
1/\tau^5)$ or $\nu$ in \eqref{nu} satisfies $ Re \nu<1/2$ and $V(\tau)=O(1/\tau^3)$. If $\Phi\in \cS(M)$, 
the following facts hold for $0<\epsilon<3/2-Re\nu$.}

{\em {\bf (a)} $\Ga\Phi$ decays faster than $1/\ell^{\epsilon}$ uniformly in the angular variables.

{\bf (b)} $\partial_\ell \Ga\Phi$ decays faster than $1/\ell^{1+\epsilon}$ uniformly in the angular variables.   

{\bf (c)} $\Ga \Phi \in \cS(\scrim)$.

{\bf (d)} Particularly, referring to the  Fourier transform $\widehat{\Ga \Phi}$ with $\Phi\in \cS(M)$, it 
holds $\widehat{\Ga \Phi} \in C^\infty((\bR\setminus\{0\})\times \bS^2; \bC)$, it vanishes uniformly in the 
angles faster than every negative power of $k$ as $|k|\to +\infty$, 
\beq\label{aggiunta}
|\partial^n_k\widehat{\Gamma\Phi}(k,\omega)| \leq C_{\delta,n}/|k|^{Re\nu+n-1/2}\:,\quad \mbox{for $0<|k|
\leq \delta$,}
\eeq
if $n=0,1,2,3$ and for some $C_{\delta,n}, 
\delta >0$.}\\

\proof{Let us start with (b) taking (\ref{proj}) into account as well as (3) in remark \ref{remarkk}. Without
losing generality, let us assume that $\ell$ is positive then
$$
|\ell^{1+\epsilon}\pa_\ell\left(\Gamma\Phi\right)(\ell,\omega)|
\leq \sup_{\omega' \in \bS^2} |\ell|^\epsilon
\left|\int_{-\infty}^{+\infty}\sp\sp\sp dk\;  e^{-i k \ell} \frac{1+|k\ell|}
{1+|k \ell|}  
\partial_k\left[ik\sqrt{\frac{kH}{2}}\widetilde\Phi(Hk,\eta(\omega'))\right]
 \right|\;.
$$
The right-hand side could in principle diverge, but we are going to show that, indeed, this is not the case.
Taking $\Psi(k,\omega)$ as a shorter notation for $\pa_k \aq i k \sqrt{\frac{kH}{2}}\widetilde\Phi(Hk,\eta(
\omega))\cq$ we get
$$
|\ell^{1+\epsilon}\pa_\ell\left(\Gamma\Phi\right)(\ell,\omega)|
\leq 
\sup_{\bS^2} |\ell|^\epsilon
\left|\int_{-\infty}^{+\infty}\sp\sp\sp dk\;   \frac{e^{-i k \ell}+i|k|\pa_k e^{-i k \ell}}
{1+|k \ell|}  
\Psi
 \right|\;.$$
Integrating by parts using \nref{uniform},
the preceding expression can be rewritten as:
$$
|\ell^{1+\epsilon}\pa_\ell\left(\Gamma\Phi\right)(\ell,\omega)|
\leq \sup_{\bS^2} |\ell|^\epsilon
\left|\int_{-\infty}^{+\infty}\sp\sp\sp dk\;  
\frac{e^{-ik\ell}}{1+|k\ell|} \aq \at  1-   i\at \frac{\sigma(k) }{1+|k\ell|}\ct  \ct \Psi  -i|k| \pa_k \Psi    
\cq 
\right|\;,
$$
where $\sigma(k)=1$ for $k\geq 0$ or $-1$ otherwise.
Now let us assume $\epsilon<1$. With this choice $|k\ell|^\epsilon/(1+|k\ell|) \leq 1$, so that we obtain the
following estimate: 
\beq\label{stimazza}
|\ell^{1+\epsilon}\pa_\ell\left(\Gamma\Phi\right)(\ell,\omega)|
\leq \sup_{\bS^2}
\int_{-\infty}^{+\infty}\sp\sp\sp dk\;  
\frac{1}{|k| ^{\epsilon}} 
\aq
2\left|{\Psi}\right|
+
\left|k\partial_k{\Psi}
\right|
\cq
\eeq
which is meaningful because the right-hand side is finite and it does not depend on the angles
since both the functions in the integral can be bounded by  $L^1(\bR,|k|^{-\epsilon} dk)$ functions 
(independent form angles), in view of (\ref{uniform1}) and (\ref{uniform}).
We have established the $(\theta,\varphi)$-uniform bound:
$$
|\pa_\ell\left(\Gamma\Phi\right)(\ell,\theta,\varphi)|\leq
\frac{C}{|\ell|^{1+\epsilon}}
$$
for some constant $C>0$.
If $\epsilon \geq 1$, one starts with the inequality
$$
|\ell^{1+\epsilon}\pa_\ell\left(\Gamma\Phi\right)(\ell,\omega)|
\leq \sup_{\omega' \in \bS^2} |\ell|^{\epsilon-1}
\left|\int_{-\infty}^{+\infty}\sp\sp\sp dk\;  e^{-i k \ell} \frac{1+|k\ell|}
{1+|k \ell|}  
\partial^2_k\left[ik\sqrt{\frac{kH}{2}}\widetilde\Phi(Hk,\eta(\omega'))\right]
 \right|\;;
$$
then the proof goes on as before since $\epsilon-1 <1$.
(a) can be similarly proved. (c) We have obtained that $\Ga\Phi$ is bounded (it being everywhere continuous 
and vanishing at infinity uniformly in the angles). Similarly, in view of Fubini-Tonelli theorem and on the 
fact that $\bS^2$ has finite measure, $\partial_\ell\Ga\Phi \in L^1(\bR\times \bS^2; d\ell dS^2)$ since 
it is continuous (so that it is bounded on compact sets $[-L,L]\times \bS^2$ 
 and decays faster than $1/\ell^{1+\epsilon}$ uniformly in the angles outside $[-L,L]$.
 As the requirement $||k \widehat{\Ga\Phi }||_\infty < \infty$ is trivially fulfilled by (3)
in remark \ref{remarkk}, 
to conclude the proof of (c) it is enough to establish that the Fourier transform of $\Gamma\Phi$ belongs 
to $L^1(\bR_+\times\bS^2, dkdS^2(\theta,\varphi))$.   Since, as we have
stated in (3) of remark \ref{remarkk}, $\widetilde\Phi$ is rapidly decreasing at infinity as a function of $k$, the 
functions ${k}^{1/2}\; \widetilde\Phi$ which is proportional to 
the Fourier transform of $\Gamma\Phi$ belongs 
to $L^1(\bR_+\times\bS^2, dkdS^2(\theta,\varphi))$ in our hypotheses. We conclude  that $\Ga \Phi \in \cS(\scrim)$.\\
The statement (d) follows immediately from (\ref{proj}) and from the estimates  in (3) of remark \ref{remarkk}.}

\remark It is worth noticing that the result was achieved thanks to the very definition of $\cS(\scrim)$.
With the stricter definition of $\cS(\scrim)$ adopted in \cite{DMP2}, where the functions are required to be 
$L^2(\bR\times \bS^2, d\ell dS^2)$ together with their $\ell$-derivative, the above result would have been 
much more  difficult, if not impossible, to establish. \\

We are now in place to state the theorem which establishes that $\Gamma$ individuates a symplectomorphism. 
This entails the identification between the bulk algebra of observables and a subalgebra of the boundary
counterpart, hence extending one of the main achievements of \cite{DMP2}, namely the theorem 4.4, to the 
physically relevant scenario $Re\nu<3/2$.

\teorema\label{projb}{Assume that $\xi$ 
and $m$ are such that  $\nu$ in \eqref{nu} satisfies $ Re \nu<3/2$, and that $V(\tau)=O(1/\tau^5)$ or
$\nu$ in \eqref{nu} satisfies $ Re \nu<1/2$, and that $V(\tau)=O(1/\tau^3)$. The following holds.}

{\em {\bf (a)} The linear map $-H^{-1} \Gamma:\mathcal{S}(M)\to C^\infty(\Im^-; \bR)$ is a symplectomorphism:
$$\sigma_{\scrim}(-H^{-1}\Gamma\Phi,-H^{-1}\Gamma\Phi')= \sigma_M(\Phi,\Phi')\:, \quad\forall\Phi,\Phi'\in
\mathcal{S}(M)$$
and $-H^{-1} \Gamma$ is injective.

{\bf (b)} There is an isometric $*$-homomorphism 
$$\imath : \cW(M) \to \cW(\scrim)\:,$$
which identifies the Weyl algebra of the observables $\cW(M)$ of the bulk with a sub $C^*$-algebra of 
$\cW(\scrim)$, and $\imath$ is completely individuated by the requirement
$$\imath\left(W_M(\Phi) \right) = W_{\scrim}(-H^{-1}\Gamma \Phi). \quad \forall \Phi \in \cS(M)$$}\\

\noindent {\em Proof}. (a) By direct inspection, if $\Phi,\Phi' \in \cS(M)$ and making use of (\ref{general})
 one almost immediately gets 
$$-2Im \int\limits_{\bR_+\times\bS^2}\;\overline
{\widetilde\Phi(k,\theta,\varphi)}\widetilde\Phi'(k,\theta,\varphi) \:k^2dk\; dS^2(\theta,\varphi)
=\sigma_M(\Phi,\Phi')\:,$$
where the integral makes sense because $k \widetilde\Phi$ and $k \widetilde\Phi'$ are elements of 
$L^2(\bR \times \bS^2, dk dS^2)$ as follows form (d) in Proposition \ref{decay}. On the other hand, we 
shall show at the end of the proof of proposition \ref{oneparticle} (and such proof does not depend on this 
one) that
\beq\label{rest}
\sigma_{\scrim}(\Gamma\Phi,\Gamma\Phi')=-2H^2 Im \int\limits_{\bR_+\times\bS^2}\;\overline
{\widetilde\Phi(k,\theta,\varphi)}\widetilde\Phi'(k,\theta,\varphi) \:k^2dk\; dS^2(\theta,\varphi)\:.
\eeq
This concludes the proof of (a) by comparison with the identity achieved above and noticing that the found symplectomorphism is injective because $\sigma_M$ is nondegenerate. 
(b) This fact straightforwardly  follows from the existence of the symplectomorphism $-H^{-1}\Gamma$ and known theorems 
on Weyl algebras \cite{BR2}. $\Box$

\ssa{Preferred state and its pullback on $M$}
The existence of the isometric $*$-homomorphism 
$\imath : \cW(M) \to \cW(\scrim)$ allows one to induce states $\omega_M$ on $\cW(M)$ from states $\omega_
\scrim$ on $\cW(\scrim)$ exploiting the pull back:
\beq\label{induction}
\omega_M (a) \doteq  \omega_\scrim\left(\imath(a) \right)\:,\quad \forall a \in \cW(M).
\eeq
The most distinguished property displayed by the quantum theory on the null surface $\scrim$ (and this is 
also true for the theory in any null infinity of asymptotically flat spacetime) is the
following \cite{DMP, DMP2}. It is possible to select a preferred algebraic quasifree state $\lambda$ on $\cW(
\scrim)$, which turns out to be invariant under the action of the conformal Killing vector $\partial_\ell$ 
and it has positive energy with respect to the self-adjoint generator of those displacements in its GNS 
representation. These features uniquely individuate the state \cite{M1,DMP2}. In the general context studied 
in \cite{DMP2}, one sees that $\lambda$ is invariant under the whole, infinite dimensional, group $SG_{\scrim
}$ of symmetries of $\scrim$ (see remark \ref{nic}) and this property is valid, referring to the BMS group, 
for the analogue state defined on the null boundary of asymptotically flat spacetimes \cite{DMP}. The state 
$\lambda$ is universal and does not depend on the particular spacetime $M$ admitting $\scrim$ as past 
boundary. Using (\ref{induction}), $\lambda$ induces a preferred state $\lambda_M$ in every
spacetime of the class under consideration. The very peculiar properties of those states were investigated in 
\cite{M2} for asymptotically flat spacetimes and in \cite{DMP2} for expanding universes. It was shown that 
$\lambda_M$ is invariant under all the isometries of $M$ (which preserve structure of $\scrim$ in the case of 
expanding universes), it has positive energy with respect to every timelike Killing vector of $M$ (which 
preserve structure of $\scrim$ for expanding universes) and, furthermore, it reduces to well-known physically 
meaningful states in the simplest cases (Minkowski vacuum and Bunch-Davies, respectively). In the case of 
asymptotically flat spacetimes $\lambda_M$ was proved to be Hadamard \cite{M2} and, thus, it can be employed 
in perturbative approaches. This is the issue we wish to examine here for our class of spacetimes.

We want now to define the preferred quasifree state $\lambda$ on $\cW(\scrim)$. The definition needs more care than 
in \cite{DMP}, since the symplectic space has been changed.
Following \cite{KW}, a quasifree state $\omega$ over a Weyl algebra $\cW(\cS)$ over the symplectic space $(\cS,\sigma)$ is individuated by its 
{\bf one-particle structure}, that is a pair $(K,H)$, where $H$ is the one-particle (complex) 
Hilbert space and $K: \cS \to H$ is an $\bR$-linear map such that (i) 
$\sigma(\psi,\psi')= -2Im \langle K\psi, K\psi'\rangle$ for all $\psi,\psi' \in \cS$ and (ii)
$\overline{K(\cS)+ i K(\cS)} = H$, the bar denoting the closure. The quasifree state $\omega$ uniquely (up to unitary transformations) 
associated with $(K,H)$ is then completely
individuated by the requirement (which extends to the whole $\cW$ by linearity and continuity)
\beq
\omega\left(W(\psi)\right) = e^{-\frac{1}{2} Re \langle K\psi, K\psi\rangle}\:, \quad \forall \psi \in \cS\:.
\eeq
The state $\omega$ turns out to be pure ({\it i.e.}, its GNS representation is irreducible) if and only if 
$\overline{K(\cS)} = H$. The GNS representation of a quasifree state $\omega$ is always a standard Fock representation with $H$ as one-
particle space, the cyclic vector is the vacuum and the representation itself maps $W(\psi)$ into $e^{
\overline{i\hat{\Phi}(\psi)}}$ where $\hat{\Phi}$ is the densely defined field operator constructed out of 
the creation and annihilation operators (see \cite{KW,Wald2,BR2} for details).

Let us come to the preferred state $\lambda$ on $\cW(\scrim)$. 
Following \cite{MP4,DMP,DMP2}, its one-particle structure $(K_\lambda,H_\lambda)$ should be made as follows.
$H_\lambda = L^2(\bR_+\times \bS^2, 2dkdS^2)$ and  $K_\lambda : \cS(\scrim) \to H_\lambda$ associates
$\psi\in \cS(\scrim)$ with its Fourier transform $\widehat{\psi}= \widehat{\psi}(k,\omega)$ restricted to the values $k \in \bR_+$. 
Differently from \cite{MP4,DMP,DMP2} where the well-posedness of the construction were guaranteed by the very 
definition of $\cS(\scrim)$ whose elements were functions of $L^2(\bR \times\bS^2; d\ell dS^2)$ with 
$\ell$-derivative in $L^2(\bR \times\bS^2; d\ell dS^2)$, now the Fourier transform has to be interpreted in 
the distributional sense rather than a Fourier-Plancherel transform. 
In principle there is no automatic reason because, with the given definition of $\cS(\scrim)$, 
the restriction of $\widehat{\psi}$ to $\bR_+$ belongs to $L^2(\bR_+\times \bS^2, 2dkdS^2)$ if
$\psi\in\cS(\scrim)$ nor for the condition (i) above stated to be valid. Therefore a result on the 
well-posedness of the construction is necessary.

\proposizione{\label{oneparticle}
Let us define
\beq\label{HL} H_\lambda \doteq L^2(\bR_+\times \bS^2, 2dkdS^2)\quad \mbox{and}\quad
K_\lambda : \cS(\scrim) \ni \psi \mapsto \Theta\cdot \widehat{\psi} \in H_\lambda\:,
\eeq 
 where $\widehat{\psi}= \widehat{\psi}(k,\omega)$ is the Fourier transform of
$\psi\in \cS(\scrim)$ and $\Theta(k)\doteq 0$ for $k\leq 0$ and $\Theta(k) \doteq 1$ otherwise.
There is a quasifree pure state $\lambda: \cW(\scrim) \to \bC$ 
whose one-particle structure is, up to unitary maps, $(K_\lambda,H_\lambda)$. More precisely:}

{\em {\bf (a)}  the $\bR$-linear map  $K_\lambda: \cS(\Im^-) \to H_\lambda$  is well defined,

{\bf (b)} $\overline{K_\lambda \left(\cS(\scrim)\right)}= H_\lambda$, the bar denoting the closure,

{\bf (c)} $\sigma_\scrim(\psi,\psi')=  -2 Im \langle K_\lambda\psi,K_\lambda\psi'\rangle$.}\\

\noindent {\em Proof}. The first statement is consequence of (a), (b) and (c). Let us prove them. (a)
If $\psi,\psi' \in \cS(\scrim)$, in view of the definition (\ref{SS}) of $\cS(\scrim)$ one has that
$$
\int_{\bR_+ \times \bS^2} 
\left|\overline{K_\lambda\psi(k,\theta,\varphi)}\; K_\lambda\psi'(k,\omega)\right| \;2k\; dk dS^2,$$
is bounded by
$$
\int_{\bR \times \bS^2} 
\left|\overline{\widehat{\psi}(k,\omega)} \widehat{\psi'}(k,\omega)\right|  \; 2k\; dk dS^2(\omega)
\leq \sup_{\omega \in \bS^2}| k\; \widehat{\psi}(k,\omega)  |  \;\int_{\bR\times \bS^2}  2\left| \widehat{\psi'}(k,\omega)\right| dk dS^2(\omega)< +\infty\:.
$$
(b) The statement is true because $K_\lambda(\cS(\scrim))$ includes the set, dense in $L^2(\bR_+\times \bS^2, 2dkdS^2)$, of the complex smooth function with compact support which do not intersect
a neighbourhood (depending of the function) of the set $\{k=0, \omega \in \bS^2\}$. Indeed, if $\phi_0$ is one of 
such functions, it can be smoothly extended in the region $k<0$ as $\phi_0(-k,\omega) \doteq \overline{\phi_0(k,\omega)}$.
The resulting function has inverse Fourier transform given by a real element of $\mS(\bR\times \bS^2)$ and thus 
it belongs to $\cS(\scrim)$. \\
(c) Let $\psi \in \cS(\scrim)$ and $\widehat{\varphi_n}\in C_0^\infty(\bR\times \bS^2; \bC)$
so that $\varphi_n(\cdot, \omega) \in \mS(\bR)$ and $\partial_\ell\psi(\cdot,\omega) \in \mS'(\bR)$
for every  $\omega\in \bS^2$. 
By standard properties of Fourier transform of Schwartz distributions, one has:
\beq \label{trick}\int_{\bR}  \varphi_n(\ell,\omega)\partial_\ell\psi(\ell,\omega) d\ell 
= -i\int_{\bR} \overline{\widehat{\varphi_n}(k,\omega)} \widehat{\psi}(k,\omega)kdk \:.\eeq
However both the right-hand side and the left-hand side can be interpreted as standard integrals, 
in our hypotheses on $\cS(\scrim)$. Now fix $\psi'\in \cS(\scrim)$ and, 
taking the angles $\omega=(\theta,\varphi)$ fixed again, 
consider  a sequence of compactly supported smooth functions $\widehat{\varphi_n}(k)$ which converges to $\widehat{\psi'}(\cdot,\theta,\varphi)$  in $L^1(\bR,dk)$. Notice that this implies that 
$||\varphi_n(\cdot,\omega)- \psi(\cdot,\omega)||_\infty \to 0$, by standard properties of Fourier transform.
As a consequence one has from (\ref{trick}):
\begin{gather*}
\int_{\bR}  \psi'(\ell,\omega)\partial_\ell\psi(\ell,\omega) d\ell  = \lim_{n\to +\infty}\int_{\bR} 
\varphi_n(\ell,\omega)\partial_\ell\psi(\ell,\omega) d\ell=\\
= \lim_{n\to+\infty}\int_{\bR} \sp -i\overline{\widehat{\varphi_n}(k,\omega)} \widehat{\psi}(k,\omega)kdk
= \int_{\bR} \sp -i\overline{\widehat{\psi'}(k,\omega)} \widehat{\psi}(k,\omega)kdk.
\end{gather*}
Concerning the third and the first  identity we have exploited the inequalities:
\begin{align}
&\int_\bR
| \overline{\widehat{\varphi_n}}\widehat{\psi}
-\overline{\widehat{\psi'}}\widehat{\psi}|\; 2k\; dk \;\leq \; 
\| 2k \; {\widehat{\psi}(\cdot,\omega)}\;\|_\infty
\| \widehat{\varphi_n}(\cdot,\omega)
-\widehat{\psi'}(\cdot,\omega)\|_{L^1} \to 0\quad \mbox{as $n\to +\infty$,}\nonumber \\
&\int_\bR 
|\varphi_n\pa_\ell {\psi}- \psi'\pa_\ell{\psi} |\; d \ell \;
\leq \|\partial_l\psi(\cdot,\omega) \|_{L^1} \|\varphi_n(\cdot,\omega) - \psi'(\cdot,\omega) \|_{\infty} \to 0
\quad \mbox{as $n\to +\infty$.} \nonumber
\end{align}
We have obtained that
\beq \label{t2}\int_{\bR}  \psi'(\ell,\omega)\partial_\ell\psi(\ell,\omega) d\ell  
= -i\int_{\bR} \overline{\widehat{\psi'}(k,\omega)} \widehat{\psi}(k,\omega)kdk\:.\eeq
We know by the proof of (a) that $\overline{\widehat{\psi'}} \widehat{\psi}\in L^2(\bR\times \bS^2, dk dS^2)$
and, by the very definition of $\cS(\scrim)$ 
it results that $\psi'\partial_\ell \psi \in L^2(\bR\times \bS^2, d\ell dS^2)$. The direct application of 
Fubini-Tonelli theorem to (\ref{t2}) yields
\beq \label{t3}\int_{\bR\times \bS^2}  \psi'(\ell,\omega)\partial_\ell\psi(\ell,\omega) d\ell dS^2  
= -i\int_{\bR\times \bS^2} \overline{\widehat{\psi'}(k,\omega)} \widehat{\psi}(k,\omega)kdk dS^2\:.\eeq
Using the fact that $\widehat{\psi}(-k,\omega) = \overline{\widehat{\psi}(k,\omega)}$
and $\widehat{\psi'}(-k,\omega) = \overline{\widehat{\psi'}(k,\omega)}$ because $\psi,\psi'$ are real, (\ref{t3})
together with the definition of $K_\lambda$ and $\sigma_{\scrim}$ implies that $\sigma_\scrim(\psi,\psi')= -2
Im \langle K_\lambda\psi,K_\lambda\psi'\rangle$, as wanted. 
This identity implies the validity of (\ref{rest}) using the fact that, from (\ref{proj}),
$$K_\lambda\left(-H^{-1}\Gamma(\Phi)\right) = -i H e^{-i\pi/4} \sqrt{\frac{Hk}{2}} \widetilde{\Phi}(Hk, \eta(
\omega))\:,$$
and taking into account that $-H^{-1}\Gamma(\Phi) \in \cS(\scrim)$ when $\Phi \in \cS(M)$ as proved in 
(a) of theorem \ref{projb}. $\Box$\\

\noindent We have now all the ingredients to construct a bulk state starting from $\lambda$, the boundary 
counterpart, proceeding as indicate in (\ref{induction}) at the beginning of this section. We define the 
quasifree state $\lambda_M$ on $\cW(M)$ individuated by the requirement:
\beq\label{state}
\lambda_M(a)\doteq\lambda(\imath(a)) \:, \quad\forall a\in\cW(M)
\eeq
It is worth stressing that in \cite{DMP2} a different definition of $\sS(\scrim)$ was exploited, however, 
as it can be checked 
by direct inspection, all the above-mentioned properties of $\lambda$ and $\lambda_M$ 
can be proved with the definition given in this paper for the whole class of spacetimes 
(which are not homogeneous nor isotropous in general) discussed i9n \cite{DMP2},
 essentially because $\lambda$ is defined employing  (\ref{HL}) also in \cite{DMP2} and because 
 the the image of the symplectomorphism which associates a wavefunction in the bulk with its restriction to $\scrim$
is included in $\sS(\scri)$ no matter which of the two definition is adopted.  
  In particular,
if $M$ is the de Sitter spacetime, $\lambda_M$ is nothing but the  Bunch-Davies vacuum
\cite{SS, BD, Allen, Kirsten} as discussed in \cite{DMP2}.

\se{On the Hadamard property.}
In this section we shall prove the main statement of the paper: \eqref{state} is of Hadamard form for every 
FRW spacetime in the class individuated by the metric (\ref{FRWconf1}) with the constraints (\ref{const})
and for values of $\nu$ in (\ref{nu}) either such that $Re \nu <1/2$ or such that $Re\nu <3/2$ though
requiring the shape of the scale factor to be that of (\ref{const2}). Let us remember that this latter request
entails the potential $V$, appearing in (\ref{rhokeq}), to satisfy $V(\tau) = O(1/\tau^5)$. 

As we stressed in the introduction, this scenario is of certain physical relevance if we think of 
inflation models where a scalar field with $\nu$ close or equal to $\frac{3}{2}$ is employed as the building 
block; remarkably the perturbative fluctuations lead to an almost homogeneous power spectrum which can be 
indirectly observed by experiments and a byproduct of the results of this section is to provide a
mathematical consistency to the underlying employed quantisation scheme. 

There are many reasons to consider Hadamard states as the most physically relevant ones and it exists a
well-developed literature discussing them, especially in relation with the problem both of construction of 
Wick polynomials and, more generally, of renormalisation in curved spacetime \cite{Brunetti,BF00,Hollands2,
BFV}.
Roughly speaking, the Hadamard property  for a state is very important in QFT in curved spacetime because it
assures  that the stress energy tensor operator \cite{Mstress,HWstress} evaluated on that
 state is renormalizable \cite{Wald2} and, 
thus, the theory might be gravitationally stable at least at perturbative level. 

We shall quickly recall the main features of the notion of an Hadamard state leaving a
reader interested in more details to specific papers \cite{KW,Radzikowski} (see also \cite{Sanders} for some recent achievements). Consider a smooth globally 
hyperbolic spacetime $(M,g)$, let $(\cS(M),\sigma_M)$ the real symplectic space of the real smooth solutions 
of Klein-Gordon equation with compactly supported Cauchy data, as defined previously. 
 $E \doteq A-R : C_0^\infty(M; \bR) \to \cS(M)$ the {\em causal propagator} \cite{KW,Wald2,BGP} associated with 
 the Klein-Gordon operator $P$ in (\ref{wave}). $A$ and $R$ are, respectively, the advanced and retarded fundamental solutions. $E$ it is known 
to be onto $\cS(M)$, with kernel given by the functions $Pf$,
for all $f\in C_0^\infty(M; \bR)$
 and it is continuous as an operator from $C_0^\infty(M; \bR)$ to $C^\infty(M; \bR)$ in the relevant topologies 
of the considered spaces of test functions.
The {\bf two-point function} of a quasifree state $\omega$ over $\cW(M)$ with one particle structure $(K,H)$
can be defined (see \cite{KW} for further details) as the quadratic form:
\beq
\omega(f,g) := \langle K(Ef), K(Eg) \rangle_H\:, \quad \forall f,g \in C_0^\infty(M; \bR)\:.
\eeq
If $(f,g) \mapsto \omega(f,g)$ is weakly continuous in each argument separately, the Schwartz kernel theorem 
assures that $\omega(\cdot,\cdot)$ uniquely individuates a distribution $\omega(\cdot)\in \mathcal{D}'(M\times M)$,
known as the {\em Schwartz kernel} of $\omega(\cdot,\cdot)$,  by requiring $\omega(f\otimes g) = \omega(f,g)$ for all $f,g \in C_0^\infty(M; \bC)$.
In fact, $\omega(\cdot)$ is nothing but the integral kernel $\omega(x,y)$ in a distributional sense:
$$\omega(f,g)=\sp\sp \int\limits_{M\times M}\sp\sp \omega(x,y)f(x)g(y)\:\: d\mu_g(x)d\mu_g(y),\quad
\omega(h)=\sp \sp \int\limits_{M\times M}\sp \sp \omega(x,y)h(x,y)\:\: d\mu_g(x)d\mu_g(y)\:, $$
if $f,g\in
C^\infty_0(M; \bC)$ and $h\in C_0^\infty(M\times M; \bC))$ and 
$d\mu_g$ denoting the metric-induced measure on $M$. In the following we shall use the same symbol to denote a quasifree state,
the associated quadratic form and its Schwartz kernel when the meaning of the symbol will be clear from the context.\\
A quasifree state $\omega$ over $\cW(M)$
is Hadamard if its kernel $\omega(x,y)$ enjoys a very peculiar behaviour at short distance of the arguments. 
We shall not enter into details here \cite{KW} because we shall deal with the {\em microlocal 
characterisation} of Hadamard states due to Radzikowski \cite{Radzikowski, Radzikowski2}.

\proposizione\label{Had}{In a globally hyperbolic spacetime $(M,g)$, consider a quasi-free state $\omega$
for the real smooth Klein Gordon field. Assume that the two-point function
of the state individuates a distribution in $\mathcal{D}'(M\times M)$. The state $\omega$ is Hadamard 
if and only if the wavefront set $WF(\omega)$ of the Schwartz kernel of the two-point function  has the form:
$$WF(\omega)=\left\{((x,k_x),(y,-k_y))\in \left(T^*M\right)^2\setminus 0\;|\; (x,k_x)\sim(y,k_y),k_x
\triangleright 0\right\}, $$
where $(x,k_x)\sim(y,k_y)$ means that it exists a null geodesic connecting $x$ and $y$ with cotangent vectors
respectively $k_x$ and $k_y$, whereas $k_x\triangleright 0$ means the $k_x$ is causal and future-directed.
Here $0$ is the zero section in the cotangent bundle.}\\

\noindent In \cite{M2}, it was proved that the analogue of the state $\lambda_M$ introduced in \cite{DMP} for 
asymptotically flat spacetimes at null infinity is Hadamard. A similar proof can be found in \cite{Hollands}
in a very different physical context. The main goal of this paper is to prove the Hadamard property for the 
states $\lambda_M$ in the considered class of FRW expanding universes admitting a past cosmological horizon. 
We state this result formally. The proof will take all the remaining part of this section and it will be 
divided in several technical steps.

\teorema\label{Ultima}{Consider a FRW spacetime $(M,g_{FRW})$ 
in the class individuated by the metric (\ref{FRWconf1}) with the constraints (\ref{const})
and for values of $\nu$ in (\ref{nu}) fulfilling either that $Re \nu <1/2$ or that  $Re \nu <3/2$ though
under the assumption (\ref{const2}).
The quasifree state $\la_M$ defined on $\cW(M)$  in \eqref{state} is Hadamard since its two-point function individuates a 
distribution of $\mathcal{D}'(M\times M)$ with wavefront set:
\beq\label{wavef}
WF(\lambda_M)=\left\{((x,k_x),(y,-k_y))\in \left(T^*M\right)^2\setminus 0\;|\;
(x,k_x)\sim(y,k_y),k_x\triangleright 0\right\}.
\eeq}

\noindent The proof of theorem \ref{Ultima} will be the topic of the rest of the paper.

\ssa{The two-point function of $\lambda_M$ individuates a distribution in $\mathcal{D}'(M\times M)$}

\vskip .2cm
\noindent As the title itself suggests, we shall now dwell into the first part of the proof of Th.\ref{Ultima}. To
start with we need the following proposition:

\proposizione\label{teo1}{With the hypotheses of Theorem \ref{Ultima}, the following facts hold for the 
two-point function $\lambda_M(f,g)$.}

{\em {\bf (a)} The two-point function of $\lambda_M$ uniquely individuates a distribution of $\mathcal{D}'(M\times M)$, this is the Schwartz
 kernel
associated with the quadratic form:
\beq \label{integstat0}\lambda_M(f,g)=\int\limits_{\bR\times\bS^2}\; 2k\;\Theta(k)\;\overline{\widehat\psi_f(k,\theta,\varphi)}
\widehat{\psi}_g(k,\theta,\varphi)dk\;dS^2(\theta,\varphi)\:,
\eeq
where $\Theta(k)\doteq 0$ if $k\leq 0$ and $\Theta(k) \doteq 0$ otherwise, $\widehat{\psi_h}$ is the $\ell$-Fourier transform of $\psi_h\doteq -H^{-1}\Ga (E h)$, for every $h \in C_0^\infty(M; \bR)$, where $\Gamma$ is defined as in 
\eqref{proje} and with $E: C^\infty_0(M; \bR) \to \cS(M)$ denoting the causal propagator of the Klein-Gordon 
operator $P$ in (\ref{wave}).

 {\bf (b)} Referring to the frame $(\ell,\theta,\varphi)$ on $\Im^-$, if $Re \nu <1$, it holds:
\beq\label{integstat}
\lambda_M(f,g)=\lim\limits_{\epsilon\to 0^+}-\frac{1}{\pi}\int\limits_{\bR^2\times\bS^2}\frac{\psi_f(\ell,
\theta,\varphi)\psi_g(\ell',\theta,\varphi)}{(\ell-\ell'-i\epsilon)^2}d\ell d\ell'dS^2(\theta,\varphi).
\eeq}

\noindent{\em Proof of Theorem \ref{teo1}}.
(a) Let us consider two solutions
$\phi_f=Ef$ and $\phi_g=Eg$ of \eqref{wave} associated with
any two functions $f,g\in C^\infty_0(M; \bR)$. Define $\psi_f\doteq -H^{-1}\Gamma\phi_f$
and $\psi_g\doteq -H^{-1}\Gamma\phi_g$. In view of Theorem \ref{oneparticle} and the definition of the state 
$\lambda_M$, we have:
$$\lambda_M(f,g)=\int\limits_{\bR\times\bS^2}\; 2k\;\Theta(k)\;\overline{\widehat\psi_f}(k,\theta,\varphi)
\widehat{\psi}_g(k,\theta,\varphi)dk\;dS^2(\theta,\varphi)\:,
$$
where $\Theta(k)\doteq 0$ if $k\leq 0$ and $\Theta(k) \doteq 0$ otherwise.
We postpone the proof of (a) at the end of the proof of the statement (b).\\
(b) Let us show that  \nref{integstat0} is equivalent to \nref{integstat} if $Re \nu<1$.
At fixed angles $\omega= (\theta,\varphi)$, we consider a sequence of real compactly-supported smooth 
functions $\varphi_n$ whose $\ell$-Fourier transform $\widehat{\varphi_n}$ converge to $\widehat{\psi_g}$ in 
the $L^1(\bR, dk)$ norm. Since $k\widehat{\psi_f}\in L^1(\bR, kdk)$ by definition of $\cS(\scrim)$ and the 
$\widehat{\varphi_n}$ are bounded, we obtain via Lebesgue's dominated convergence:
\begin{gather*}
\int_{\bR} 2k\;\Theta(k)\;\overline{\widehat\psi_f}(k,\omega)\widehat{\varphi_n}(k,\omega) dk
=
\lim_{\epsilon\to 0^+} \int_0^\infty e^{-\epsilon k } \;2k\; \overline{\widehat{\psi_f}}(k,\omega)\; 
\widehat{\varphi_n}(k,\omega) \:dk 
\end{gather*}
Now notice that both $k \mapsto 2k e^{-\epsilon k}\Theta(k)\widehat\psi_f(k,\omega)$ and 
$k \mapsto \widehat{\varphi_n}(k,\omega)$ are functions of $L^2(\bR,dk)$, the former because 
of (d) in Proposition {\ref{decay}}. Therefore the Fourier transform
can be interpreted as the Fourier-Plancherel one - say $\mF$ - and
\beq\label{ftrans}
\langle 2k e^{-\epsilon k}\Theta\widehat\psi_f, \widehat{\varphi}_n\rangle_{L^2(\bR, dk)}=\left\langle {
\mF}^{-1}\left(2k\Theta e^{-\epsilon k}\widehat\psi_g\right), \varphi_n\right\rangle_{L^2(\bR, dk)}, 
\eeq
We can now use the convolution theorem in $L^2(\bR,dk)$ to rearrange the 
right-hand side of the internal product as:
$$
\mathcal{F}^{-1}\left(\Theta e^{-\epsilon
k}\widehat\psi_f\right)(\ell,\omega)=\frac{1}{\pi}\int\limits_\bR\frac{\partial_{\ell'}\psi_f(\ell',
\theta,\varphi)}{(\ell-\ell'-i\epsilon)}d\ell', 
$$
With this in mind we have that:
\begin{gather*}
\langle 2k e^{-\epsilon k}\Theta\widehat\psi_f, \widehat{\varphi_n}\rangle_{L^2(\bR, dk)}=
-\frac{1}{\pi}\int\limits_{\bR}d\ell\;
\varphi_n(\ell,\omega)\int\limits_\bR d\ell'\;\frac{\psi_f(\ell',\omega)}{(\ell-\ell'-i\epsilon)^2}
\end{gather*}
where in the last equality we integrated by parts.
Using the fact that that, uniformly in the angles 
 $\omega$, $\psi_f$ is bounded and tends to
$0$ as $1/|\ell|^\delta$ with $\delta \in (0, 3/2 -Re \nu)$ (see proposition \ref{decay} where $\delta$ 
was indicated by $\epsilon$), one sees by direct computation that, for $h\equiv 1$ or $h= \varphi_n$ or $h=
\psi_g$ and where $C\geq 0$ does not depend on angles:
\beq \label{lun}\int\limits_{\bR\times\bR}d\ell d\ell'\;\left|h \frac{\psi_f(\ell',\omega)}{(\ell-\ell'-i\epsilon)^2}\right| \leq ||h(\cdot,\omega)||_\infty
\int\limits_{\bR\times\bR}d\ell d\ell'\;\left|\frac{C}{(1+ |\ell|^\delta)(\ell-\ell'-i\epsilon)^2}\right|<
+ \infty\:,\eeq
when one chooses $\delta>1/2$, and this is possible when $Re \nu< 1$. Particularly this implies that, in 
view of Fubini-Tonelli theorem, the integrals:
$$\int\limits_{\bR\times\bR}\sp d\ell d\ell'\frac{\varphi_n(\ell,\omega)\psi_f(\ell',\omega)}{(\ell-\ell'-
i\epsilon)^2}=\int\limits_{\bR}\sp d\ell\sp\int\limits_\bR \sp d\ell'\frac{\varphi_n(\ell,\omega)\psi_f(
\ell',\omega)}{(\ell-\ell'-i\epsilon)^2},$$ 
and
$$
\int\limits_{\bR\times\bR}\sp d\ell d\ell' \frac{\psi_g(\ell,\omega)\psi_f(\ell',\omega)}{(\ell-\ell'-i
\epsilon)^2}=\int\limits_{\bR}\sp d\ell\spa\int\limits_\bR \sp d\ell'\frac{\psi_g(\ell,\omega)\psi_f(\ell',
\omega)}{(\ell-\ell'-i\epsilon)^2},$$
are meaningful and the end point is
$$\langle 2k e^{-\epsilon k}\Theta\widehat\psi_f, \widehat{\varphi_n}\rangle_{L^2(\bR, dk)}=
-\frac{1}{\pi}
\int\limits_{\bR \times \bR}d \ell d\ell'\;\frac{\varphi_n(\ell,\omega)\psi_f(\ell',\omega)}{(\ell-\ell'-i\epsilon)^2}\:.
$$
On the other and, since $||(\psi_g(\cdot,\omega)-\varphi_n(\cdot,\omega))||_\infty \to 0$ as $n\to +\infty$
because $\widehat{\varphi_n}$ converge to $\widehat{\psi_g}$ in the $L^1(\bR, dk)$, for $n\to +\infty$ we have:
$$\left|\int\limits_{\bR}d\ell\;
\int\limits_\bR d\ell'\; \frac{(\psi_g-\varphi_n)\psi_f(\ell',\omega)}{(\ell-\ell'-i\epsilon)^2}\right|
\leq ||(\psi_g-\varphi_n)||_\infty
\int\limits_\bR d\ell'\;\left|\frac{\psi_f(\ell',\omega)}{(\ell-\ell'-i\epsilon)^2}\right| \to 0
$$
so that, as $k \mapsto 2k e^{-\epsilon k}\Theta(k)\widehat\psi_f(k,\omega)$ is bounded,
$$ \langle 2k e^{-\epsilon k}\Theta\widehat\psi_f, \widehat{\psi_g}\rangle_{L^2} =
\lim_{n \to +\infty} \langle 2k e^{-\epsilon k}\Theta\widehat\psi_f, \widehat{\varphi_n}\rangle_{L^2} =
-\frac{1}{\pi}\int\limits_{\bR^2}\frac{\psi_f(\ell,
\omega)\psi_g(\ell',\omega)}{(\ell-\ell'-i\epsilon)^2}d\ell d\ell'\:,$$
that is
$$\int_0^\infty e^{-\epsilon k } \;2k\; \overline{\widehat{\psi_f}(k,\omega)}\; 
\widehat{\psi_g}(k,\omega) \:dk = -\frac{1}{\pi}\int\limits_{\bR^2}\frac{\psi_f(\ell,
\omega)\psi_g(\ell',\omega)}{(\ell-\ell'-i\epsilon)^2}d\ell d\ell'\:.$$
Integrating $\omega$ over the compact set $\bS^2$ (and this is possible concerning the left-hand side 
because the integrand belong to $L^1(\bR\times \bS^2, 2kdk dS^2)$ in view of the definition of $\cS(\scrim)$, whereas 
(\ref{lun}) holds for the integrand in the right-hand side), it arises
$$\int_{\bR\times \bS^2} e^{-\epsilon k } \;2k\; \overline{\widehat{\psi_f}(k,\omega)}\; 
\widehat{\psi_g}(k,\omega) \:dkdS(\omega) = -\frac{1}{\pi}\int\limits_{\bR^2\times \bS^2}\frac{\psi_f(\ell,
\omega)\psi_g(\ell',\omega)}{(\ell-\ell'-i\epsilon)^2}d\ell d\ell'dS^2(\omega)\:.$$
Lebesgue's dominated convergence theorem produces immediately (\ref{integstat}) when (\ref{integstat0})
is assumed.\\
We conclude now the proof of the statement (a), proving that the two-point function of $\lambda_M$
individuates a distribution in $\mathcal{D}'(M\times M)$. To this end we are going to show that, for any 
fixed $f\in C_0^\infty(M;\bR)$, 
$\lambda_M(f,\cdot )$ is the weak limit of a sequence of distributions $R_{f,n} \in \mathcal{D}'(M)$ and, for
any fixed $g\in C_0^\infty(M;\bR)$, $\lambda_M(\cdot,g)$ is the weak limit of a sequence of distributions 
$L_{g,n}\in\cD'(M)$. This fact implies that $\lambda_M(f,\cdot),\lambda_M(\cdot, g) \in \mathcal{D}'(M)$
and the map $C_0^\infty(M;\bR) \ni f \mapsto \lambda_M(f,\cdot) \in  \mathcal{D}'(M)$ is well-defined
and sequentially continuous in particular. The standard argument based on Schwartz' integral kernel theorem 
finally implies that $\lambda_M(\cdot,\cdot)\in\mathcal{D}'(M\times M)$.\\
The required sequences of distributions are defined as 
$R_{f,n}(g)\doteq  \la_n(f,g)$ and $L_{g,n}(f) \doteq \la_n(f,g)$ where:
\beq\label{statiausiliari}
\la_n(f,g) \doteq \lim\limits_{\epsilon\to 0^+} \int_{\bR^2 \times \bS^2} \frac{(\psi_f\chi_n)(\ell',\theta,
\varphi)\; (\psi_g \chi_n) (\ell,\theta,\varphi)}{(\ell-\ell'-i\epsilon)^2}\; d\ell d\ell'dS^2,
\eeq
Above $\chi_n(\ell) \doteq \chi(\ell/n)$, $n=1,2,\ldots$ are some cutoff functions on $\bR\times \bS^2$ which
are constant in the angular variables; they are defined out of $\chi\in C_0^\infty(\bR; \bR)$, such that 
$\chi(0)=1$. The functionals $R_{f,n}$ and $L_{g,n}$ are distributions because $\lambda_n\in\mathcal{D}'(M
\times M)$ since:
\begin{gather}
|\la_n(f,g)| \leq C_n \sum_{|m|< N} \sup |\pa^m (\chi_n\psi_f)| \sum_{|m'|<N'}\sup |\pa^{m'}(\chi_n\psi_g)| 
\notag \\ 
\leq 
C_n \sum_{|m|< N}\sup_{supp \chi_n} |\pa^m(\chi_n\psi_f)| \sum_{|m'|<N'}\sup_{supp \chi_n} |\pa^{m'}(\chi_n
\psi_g)|\leq C'_n \sum_{|p|<M'}\sup |\pa^p f|\sum_{|p'|<M'}\sup |\pa^{p'} g|. 
\end{gather}
The first estimate holds because the kernel $1/(\ell-\ell' -i0^+)^2$ is a well-defined distribution of 
the space $\mathcal{D}'((\bR\times \bS^2)\times (\bR\times \bS^2)))$.
In the last estimate we have used the fact that 
 $\chi_n\psi_f=\chi_nE f$ where  $supp \chi_n$ is compact, and the continuity of the causal propagator $E :
C^\infty_0(M,\bR) \to C^\infty(M,\bR)$ with respect to the relevant topologies.
To conclude the proof it is sufficient to prove that $R_{f,n} \to \lambda(f,\cdot)$ and 
$L_{g,n} \to \lambda(\cdot, g)$ in weak sense, as $n\to +\infty$. To this end we notice that, exploiting the proof 
of the part (a) in the much easier  situation where $\psi_f\chi_n$ and $\psi_g\chi_n$ have compact support, one achieves:
\beq \label{integstat01} \la_n(f,g) = \int\limits_{\bR\times\bS^2}\; 2k\;\Theta(k)\;\overline{\widehat{\chi_n\psi_f}}\:
\widehat{\chi_n\psi_g} dk\;dS^2\:.\eeq
Therefore, if one performs the integrals over $\bR_+\times\bS^2$, it holds
\begin{align}
&|R_{f,n}(g) - \la_M(f,g)|= |L_{g,n}(f) - \la_M(f,g)| =
|\la_n(f,g) - \la_M(f,g)| \nonumber \\
&\leq \left|\int \aq\overline{\widehat {\chi_n\psi_f}} \widehat{\chi_n\psi_g}\; 
- \overline{\widehat {\psi_f}} \widehat{\psi_g}
\cq
2k\;dk dS^2 \right|\leq 
\int \left[\left|\overline{\widehat {\chi_n\psi_f}} \widehat{\chi_n\psi_g}\; 
- \overline{\widehat {\chi_n\psi_f}} \widehat{\psi_g}
\right|
+
\left|
\overline{\widehat {\chi_n\psi_f}} \widehat{\psi_g}\; 
- \overline{\widehat {\psi_f}} \widehat{\psi_g}
\right|
\right]\;2k dk dS^2\nonumber\\
&\leq 2||k\overline{\widehat {\chi_n\psi_f}}||_\infty
\int \left| \widehat{\chi_n\psi_g}\; 
-  \widehat{\psi_g}
\right|dk dS^2
+ 2||k\widehat{\psi_g}||_\infty\int 
\left|
\overline{\widehat {\chi_n\psi_f}} \; 
- \overline{\widehat {\psi_f}}
\right|\;dk dS^2\nonumber \:.
\end{align}
Above $k\widehat \psi_g$ is bounded by definition of $\cS(\scrim)$, and, if one makes use both of the 
convolution theorem for $L^1$ functions and of the definition of Fourier transform as well as that of $\cS(
\scrim)$, one finds by direct inspection:
$$|2k\widehat{\chi_n \psi_f}| \leq 2(2\pi)^{-1/2}||\chi ||_{\infty} ||\partial_\ell \psi ||_{\infty} + 2||\psi_f||_{L^1} 
 (2\pi)^{-1/2}
\int_\bR |\chi'(\ell /n)| d(\ell /n) < C_f <\infty\:,$$ 
where $C_f$ does not depend on $n$, though it depends on the fixed function $\chi$. We conclude that:
$$
|R_{f,n}(g) - \la_M(f,g)|= |L_{g,n}(f) - \la_M(f,g)|
\leq C_f\spa
\int \spa\left| \widehat{\chi_n\psi_g}\; 
-  \widehat{\psi_g}
\right|\;dk dS^2
+ 2 ||k\widehat{\psi_g}||_\infty \sp\int \spa
\left|
\overline{\widehat {\chi_n\psi_f}} \; 
- \overline{\widehat {\psi_f}}
\right|\;dk dS^2
$$
To conclude the proof, it is sufficient to prove that $\widehat{\chi_n\psi_f} \to \widehat \psi_f$ 
and $\widehat{\chi_n\psi_g} \to \widehat \psi_g$in $L^1(
\bR\times\bS^2, dkdS^2)$. To this end consider 
$\rho_\delta\in C_0^\infty(\bR;\bR)$ such that $\rho_\delta(0)=1$, $0\leq |\rho(k)|\leq 1$ and $supp \rho_
\delta \subset [-\delta,\delta]$, and define $\rho'_\delta \doteq 1- \rho_\delta$ which is non-negative and 
it vanishes in a neighbourhood of $k=0$. With this definition it holds, making use of the convolution theorem 
in $L^1$
\beq\label{decom} \int_{\bR \times \bS^2}  \left|\widehat{\chi_n\psi_f} - \widehat{\psi_f}\right| dk dS^2 
\leq\int_{\bR \times \bS^2}  \left|\widehat{\chi_n} * \left(\rho_\delta \widehat{\psi_f}\right) - 
\rho_\delta \widehat{\psi_f}\right| dk dS^2 +
\int_{\bR \times \bS^2}  \left|\widehat{\chi_n} * \left(\rho'_\delta \widehat{\psi_f}\right) - \rho'_\delta 
\widehat{\psi_f}\right| dk 
dS^2\:. \label{qfn}
\eeq
On the other hand, using particularly  the fact that $\widehat{\chi_n}(k)= n \widehat{\chi}(kn)$ and changing 
the coordinates in the convolution integral, one has that the first integral in the right-hand side 
is dominated by:
$$\int_{\bR \times \bS^2}  \left|\widehat{\chi_n} * \left(\rho_\delta \widehat{\psi_f}\right)\right| + \left|
\rho_\delta \widehat{\psi_f}\right| dk dS^2 
 \leq \int_{\bS^2} dS^2 \int_\bR dh |n \widehat{\chi}(nh)| \int_{-\delta}^{\delta}|\widehat{\psi_f}(p,\omega)| dp +
  \int_{\bS^2}\int_{-\delta}^{\delta} |\widehat{\psi_f}| dk dS^2\:.
$$
Noticing that $|\widehat{\psi_f}(k,\omega)| \leq B/|k|^\epsilon$ about $k=0$ as consequence of (d) in 
Proposition \ref{decay}, we achieve the final bound, for some constant $B'\geq 0$ independent from $\delta$:
\begin{gather*}
\int_{\bR \times \bS^2}  \left|\widehat{\chi_n} * \left(\rho_\delta \widehat{\psi_f}\right) - \rho_\delta 
\widehat{\psi_f}\right| dk dS^2\leq 
\int \left|\widehat{\chi_n} * \left(\rho_\delta \widehat{\psi_f}\right)\right| + \left|
\rho_\delta \widehat{\psi_f}\right| dk 
dS^2  
\leq\left(4\pi\int_\bR du | \widehat{\chi}(u)| + 1\right)B'\delta^{1-\epsilon}\:.
\end{gather*}
Concerning the second integral in the right-hand side of (\ref{qfn}) we observe that, as $n\to +\infty$:
$$\left(\widehat{\chi_n} * \left(\rho'_\delta \widehat{\psi_f}\right)\right)(k,\omega) = \int_\bR \widehat{
\chi}(p) \left(\rho'_\delta \widehat{\psi_f}\right)(k-p/n, \omega) dp \to 
\int_\bR \widehat{\chi}(p) dp \left(\rho'_\delta \widehat{\psi_f}\right)(k,\omega) = \left(\rho'_\delta 
\widehat{\psi_f}\right)(k,\omega)\:,
$$
in view of Lebesgue's dominated convergence theorem (taking into account that $\rho'_\delta \widehat{\psi_f}$
is bounded by construction -- the only singularity has been cancelled by $\rho'_\delta$, and that $\widehat{
\chi}\in L^1(\bR, dk)$); moreover, if one computes the inverse Fourier transform of $(1+k^2)\left(\widehat{\chi_n
} *\left(\rho'_\delta \widehat{\psi_f}\right)\right)$ taking into account both that the arguments of the 
convolution are Schwartz functions and that $\chi_n(\ell) = \chi(\ell/n)$, one sees that it can be bounded 
by a Schwartz function $s$ independent from $n$ and the angles, so that:
$$\left|(1+k^2)\left(\widehat{\chi_n
} *\left(\rho'_\delta \widehat{\psi_f}\right)\right)\right| \leq  \frac{1}{\sqrt{2\pi}}\int_{\bR} \left|e^{ik\ell}s(\ell)\right| d\ell \doteq K\:.$$
Therefore there is a constant $K\geq 0$ with:
\beq 
\left|\left(\widehat{\chi_n} * \left(\rho'_\delta \widehat{\psi_f}\right)\right)(k,\omega)\right|\leq\frac{K}
{1+k^2}\:, \label{bl}
\eeq
We are, thus, allowed to apply again Lebesgue's theorem to the second integral in (\ref{qfn}), concluding that it  vanishes 
for $n\to +\infty$,
$$
\int_{\bR \times \bS^2}  \left|\widehat{\chi_n} * \left(\rho'_\delta \widehat{\psi_f}\right) - \rho'_\delta 
\widehat{\psi_f}\right| dk dS^2\to 0\:.$$
Summarising and focusing back on the right-hand side of (\ref{qfn}), we can write, for every fixed $\epsilon\in (0,1)$:
\begin{gather}
0 \leq\liminf_n \int_{\bR \times \bS^2}  \left|\widehat{\chi_n\psi_f} - \widehat{\psi_f}\right| dk dS^2 
\leq\limsup_n \int_{\bR \times \bS^2}  \left|\widehat{\chi_n\psi_f} - \widehat{\psi_f}\right| dk dS^2\leq
\nonumber \\
A \delta^{1-\epsilon} \spa +\spa \limsup_n\int  \left|\widehat{\chi_n} * \left(\rho'_\delta \widehat{
\psi_f}\right) - \rho'_\delta \widehat{\psi_f}\right| dk 
dS^2 = A \delta^{1-\epsilon} \spa +\sp \spa \lim_{n\to +\infty}\int  \left|\widehat{\chi_n} * 
\left(\rho'_\delta \widehat{\psi_f}
\right) - \rho'_\delta \widehat{\psi_f}\right| dk 
dS^2
=
 A \delta^{1-\epsilon}\:, \nonumber
 \end{gather}
where the constant $A\geq 0$ does not depend on $\delta>0$ which can be taken arbitrarily small.
 This result immediately implies that $\widehat{\chi_n\psi_f} - \widehat{\psi_f}\to 0$ in the topology of $L^1(\bR\times \bS, dk
dS^2)$ and since the analogue holds for $g$, it concludes the proof of (a) and of the theorem.
$\Box$\\

\ssa{The general strategy to establish the identity \nref{wavef}}

\vskip .2cm

\noindent We can carry on with the proof of the 
theorem \ref{Ultima} proving that the wavefront set of $\lambda_M$ is that stated in (\ref{wavef}).
By construction, the distribution $\la_M \in \cD'(M\times M)$ satisfies the further properties:
$$
\la_M(f\otimes Pg) =  \la_M(Pf \otimes g) = 0  \quad {\bf (KG)}, \qquad  
\la_M(f\otimes g) - \la_M(g \otimes f) = E(f\otimes g) \quad {\bf (Comm)},
$$ 
where, in the second formula, $E$ is the Schwartz kernel of the causal propagator
 which exists in accordance with the above-mentioned continuity properties of the causal propagator.  As is well known \cite{Radzikowski,SV01,SVW02,M2}, the inclusion $\supset$ in \nref{wavef}  follows 
from $\subset$ when one applies the celebrated {\em theorem 
of propagation of singularities} due to H\"ormander \cite{Hormander}, in combination with (KG) and (Comm). So, only the inclusion
$\subset$ has to be established.
In order to prove that inclusion, we would like to interpret the  $\la_M$
as a composition of distributions, though this idea will not turn out to be truly conclusive. To this end notice that,
in view of Proposition \ref{teo1}, for $Re \nu <1$,
 the two-point function of $\la_M$ in 
\nref{integstat} reads:
\beq\label{startF}
\la_M(f,g)= T \left( (\Ga E f)\otimes (\Ga E g) \right),
\eeq
where $\Ga Ef $ represents the restriction to $\scrim$ of the wave function $Ef$, $E$ being the causal 
propagator,
and $T$ is an integral operator whose integral kernel can be thought of as the distribution in $\cD'(\scri\times \scri)$:
\beq\label{distributionT} 
T(\ell, \omega,\ell',\omega'):=-\frac{1}{H^2\pi^2 (\ell-\ell'-i0^+)^2} \otimes \delta (\omega,\omega')\:.
\eeq
Above, $H$ is the Hubble constant, $\delta (\omega,\omega')$ is the standard delta distribution on  $\bS^2$ to be integrated 
with respect to the standard measure on the unit $2$-sphere, whereas $\ell$ is the null coordinate on $\bR$ 
as in \nref{Bondiform}. 
Looking at  (\ref{startF}), our strategy  to
establish $\subset$ in \nref{wavef}, concluding the proof of Theorem \ref{Ultima}, will be the following.
 First of all we shall 
prove that, if $I$ is the identity from $C_0^\infty(M; \bC) \to C_0^\infty(M;\bC)$
(i.e. the Schwartz kernel in $\cD'(M\times M)$ given by the constant function $1$)  the product 
of distributions
\beq\label{LaK} K \doteq  (T\otimes I) (\Gamma E \otimes \Gamma E)  \in \cD'\left((\scrim\times \scrim) 
\times (M \times M)\right),\eeq
is well defined. Now notice that, if $\cK : C_0^\infty(\scrim \times \scrim ;\bC)  \to \cD'(M \times M)$
is the continuous map associated with the kernel $^tK$ in view of Schwartz kernel theorem, formally   
speaking, (\ref{startF}) would hold when:
$$\lambda_M = \cK(1 \otimes 1)\:,$$
where $1: \scrim \to \bR$ is function which takes the value $1$ constantly. However that does not make sense 
in general because, in particular, $1\not \in C_0^\infty(\scrim; \bC)$. As a matter of fact, we shall instead replace
the function $1$ with a sequence of  functions $\chi_n \in  C_0^\infty(\scrim;\bC)$ which tends to $1$ in a suitable sense
 and we shall prove that, 
in the general case $Re \nu < 3/2$  {\em so that the distribution $\lambda_M$ is defined by} (\ref{integstat0}), it holds
\beq \label{CHPT}\cK(\chi_n \otimes \chi_n) \doteq \lambda_n \to \lambda_M \:, \quad \mbox{as $n\to +\infty$\:,}\eeq
where the convergence is valid in the sense 
of H\"ormander's pseudo topology \cite{Hormander}.
We also prove that, for every $n\in \bN$, it holds:
$WF(\lambda_n)\subset \left\{((x,k_x),(y,-k_y))\in \left(T^*M\right)^2\setminus 0\;|\;
(x,k_x)\sim(y,k_y),k_x\triangleright 0\right\}$.
Taking this shape of $WF(\lambda_n)$ and (\ref{CHPT}) into account,  the properties of the notion of convergence 
in H\"ormander's pseudo topology will imply 
$\subset$ in \nref{wavef}, concluding the proof of Theorem \ref{Ultima}.

\noindent 
\sssa{Restriction to $\scrim$ of  $E$ and well-definiteness of  $K$ in (\ref{LaK})}

\vskip .2cm

\noindent We prove here that $K$ in (\ref{LaK}) is a well-defined distribution 
given by the product of two distributions. 
To this end, we need a result on the extension/restriction of one entry of the causal 
propagator $E$ to 
the horizon in order to define $\Gamma E$ appearing in (\ref{LaK})
as an element of $\cD'(\scri \times M)$. 
Next we pass to analyse $T$ and to  show that the distribution $K$ is well defined.\\
 Since both $M$ and $\widehat M$ are globally 
hyperbolic, and $M\subset\widehat M$, Eq.\eqref{wave} admits unique causal propagators, $E$ on $M$, and $\widehat E$ on $\widehat M$.
By uniqueness $\widehat E\spa \rest_{M\times M} = E$ when both are viewed as Schwartz kernels.
$\widehat E$ can be restricted
to $\Im^-$ in the left argument, when the other ranges in $M$, giving rise to a distribution in $\cD'(\scri \times M)$
we shall indicate by $\Gamma E = \Gamma E (z,x)$.  Indeed,
one has \cite{Radzikowski}:   
\beq\label{WFE}
WF(\widehat{E}):= \left\{\right.
( (z,k_z), (x, -k_x ))\in\left(T^*\widehat{M}\right)^2\setminus 0\;\left|\;( z, k_z)  \sim (x, k_x ) \right\},
\eeq
where $(x, k_x ) \sim (z, k_z)$ means that it exists a null geodesic $\ga$ from $x$ to $z$ where $k_x$ is the
cotangent vector in $x$ and $k_z$ the cotangent vector in $z$.
To restrict its left entry  to $\scrim$, we consider the immersion map $j:\scrim\times \widehat M \to\widehat M
\times\widehat M$, $j(\ell,\theta,\varphi, x)=((a= 0,\ell,\theta,\varphi),x)$ in the Bondi like coordinates 
as in \nref{Bondiform}.
According to the theorem 8.2.4 of \cite{Hormander}, that restriction is meaningful provided 
$
WF(\widehat{E})\cap N_j =\emptyset,
$
where $N_j$ is the set in $T^*\widehat M$ of normals to $\scrim$. In the case under investigation
$$
N_j:= \{((z, k_z ), (x, 0 )) \in T^*\widehat M \times  T^*\widehat M\;  |\; z
\in \scrim \: x\in \widehat M\:,  k_z = (k_z)_a da, (k_z)_a \in\bR
\}.
$$
Since null structures are preserved by conformal rescaling and since \eqref{FRWconf1} entails that the
spacetimes under consideration are conformally related to Minkowski spacetime, the null geodesics of
$(M,g_{FRW})$ have the same causal structure as those of $(\bR^4,\eta)$. If we take into account this remark 
in combination with the definition \nref{WFE} according to which each pair of points $(x,y)$ is joined by a
null geodesic, we can prove that 
$WF(\widehat{E})\cap N_j =\emptyset$ slavishly following the proof of the analogous statement in 
\cite{M2}.  So, $\Gamma E := {\widehat E}\spa\rest_{\scrim \times M}$ is well defined and
Theorem 8.2.4 of \cite{Hormander} guarantees that:
\beq \label{WFTE} WF(\Gamma E)\subset\{\left( (s,k_s) , (x, -k_x )\right)\in\left(T^*\scrim\times T^*M\right)\setminus 0\;
|\; (s, k_s ) \sim (x, k_x ), (k_s )_l \neq 0\}\:.\eeq
A bound for  $WF(\Ga E \otimes \Ga E)$ can be obtained by Theorem 8.2.9 in \cite{Hormander}, via the general formula:
\beq \label{prod}
WF(u\otimes v) \subset (WF(u)\times WF(v)) \cup ((supp \:u \times \{0\}) \times WF(v))
\cup ( WF(u) \times (supp\: v \times \{0\}))\:.
\eeq
Let us pass to analyse $T \otimes I$. 
Obviously one has:
\beq \label{AGGF} WF(T\otimes I) = WF(T)\times ((M\times M) \times \{0\})\:.\eeq
The wavefront set of $T$ was already
discussed in section 4.3 of \cite{M2} and we here summarise it:
\proposizione{ \label{PT}
$WF(T):= A\cup B$ where:
\begin{align}
&A:= \left\{ \left.\at (\ell,\om,k,\bk) , (\ell',\om',k',\bk') \ct \in \left(T^*\scrim\right)^2\setminus 0\: \right|  
 \:
 \ell=\ell', \om=\om', 0 < k = - k', \bk=-\bk' \right\},\nonumber\\
&B:= \left\{ \left.\at (\ell,\om,k,\bk) , (\ell',\om',k',\bk') \ct \in \left(T^*\scrim\right)^2\setminus 0 
\:\right|\: \om=\om',  k = k'= 0 , \bk=-\bk'  \right\}\;,\nonumber
\end{align}
with $\ell\in\bR$, $k\in T^*_\ell\bR$, $\omega\in\bS^2$, whereas $\bk\in T_\omega^*\bS^2$, and $0$ is the zero-section 
of $\left(T^*\scrim\right)^2 \equiv T^*\left(\scrim\times \scrim\right)$.
}\\

\noindent We can conclude that $K$ in (\ref{LaK}) is a well-defined distribution in 
$D'(M\times M\times\Im^-\times\Im^-)$, since the sufficient condition for the existence of 
the product of two distributions, stated
 in term of wavefront sets in theorem 8.2.10 in \cite{Hormander} is valid: there is no $(x,\xi) \in WF(T\otimes I)$
 with $(x,-\xi) \in WF(\Gamma E \otimes \Gamma E)$.
Furthermore the following inclusion holds in view of the previously cited theorem and (\ref{AGGF}):
\begin{align}\label{INCLK}
WF\left(K\right) &\subset\left\{( (s,p_s+\tilde{p}_s), (s',p_{s'}+\tilde{p}_{s'}), (x,k_{x}),(y,k_y)) \in T^*\scrim \times 
T^*\scrim \times T^*M \times T^*M \:\nonumber \right.\\
 & \:\left|\: ((s,p_s), (s',p_{s'})) \in WF(T) \:\mbox{or} \: p_s=p_{s'}=0\;,\textrm{and}\right. \nonumber \\
 & \:\:\:\left.((s,\tilde{p}_s), (s',\tilde{p}_{s'}), (x,k_{x}),(y,k_y)) \in WF\left(\Gamma E\otimes\Gamma E\right) \:\mbox{or} \: \tilde{p}_s=\tilde{p}_{s'}=k_x  =k_y =0\right\}\:.
\end{align}

\sssa{On the sequence of auxiliary distributions $\la_n$ and  their wavefront set}

\vskip .2cm

\noindent To define the sequence of distributions satisfying (\ref{CHPT}), 
let us first fix a  function $\chi\in 
C^\infty_0(\bR;\bR)$ depending on the variable $\ell$ only, and such that $\chi(0)=1$ and define 
$\chi_n \in C_0^\infty(\scri; \bR)$ as:
\beq\label{sequence}
\chi_n(\ell,\omega)\doteq\chi\left(\frac{\ell}{n}\right)\:, \quad \mbox{if $(\ell,\omega) \in \bR\times \bS^2$, }\forall n\in\mathbb{N}\;.
\eeq
Hence we can define the following sequence of distributions, which are well defined because as proved beforehand $K\in \cD'((\scri\times \scri)\times (M\times M))$,
\beq \label{LN}
\lambda_n\doteq\mathcal{K}(\chi_n\otimes \chi_n)\in\mathcal{D}'(M\times M)\:,
\eeq
where $\mathcal{K}: C_0^\infty(\scrim\times \scrim; \bC) \to \cD'(M\times M)$ is the continuous operator uniquely 
associated to $^tK$ in accordance with Schwartz kernel theorem.
These distributions have been already used  in the proof of (a) of Proposition \ref{teo1}.
The wavefront set of $\lambda_n$ satisfies the following inclusion, which can be readily inferred out of Theorem 8.2.12 in
\cite{Hormander}:
$$
WF(\lambda_n)\subset
\left\{\left((x,k_x),(y,k_y)\right)\;|\;
\left((s,0),(s',0), (x,k_x),(y,k_y)\right)\in WF(K)\:,\nonumber  \mbox{for some
} s,s' \in supp \chi_n \right\}\:.
$$
As $supp \chi_n$ becomes larger and larger as $n\to +\infty$,
if one wants to achieve a $n$-uniform bound of $WF(\lambda_n)$,
 the last requirement has be dropped and replaced with
 $s,s' \in \scrim$.
\beq\label{WF}
WF(\lambda_n)\subset
\left\{\left((x,k_x),(y,k_y)\right)\;|\;
\left((s,0),(s',0), (x,k_x),(y,k_y)\right)\in WF(K)\:, \mbox{for some
} s,s' \in \scrim \right\}.
\eeq
Taking Proposition \ref{PT} and equations (\ref{WFTE}),  (\ref{prod}) and (\ref{INCLK})  into account, with a laborious but elementary computation, this $n$-uniform estimate can be 
formally restated as: 
\proposizione{The elements $\la_n \in \cD'(M\times M)$ defined in (\ref{LN}) satisfy: 
\beq\label{NU}
WF(\lambda_n)\subset   \mathcal{V} \doteq \left\{((x,k_x),(y,-k_y))\in \left(T^*M\right)^2\setminus 0\;|\;
(x,k_x)\sim(y,k_y),k_x\triangleright 0\right\}.
\eeq
}   

\noindent Notice that  $\mathcal{V}$ is a closed subset in $\left(T^*M\right)^2\setminus 0$. As an immediate 
but indirect proof, simply notice that, in view of Radzikowski's achievements, $\mathcal{V}$ is the wavefront set
(and thus a  closed subset in $\left(T^*M\right)^2\setminus 0$ by definition of wavefront set)
of any Hadamard state on the globally hyperbolic spacetime $(M, g_{FRW})$ (and every globally hyperbolic spacetime admits
Hadamard states as is well known \cite{Wald2}).

\sssa{Proof of the fact that $\lambda_n \to \lambda_M$ in $\cD'_{\mathcal{V}}(M\times M)$, 
and the consequent $WF(\lambda_M)$} \label{seccentr}

\vskip .2cm

\noindent In order to complete the proof of Theorem \ref{Ultima},
establishing  the inclusion $\subset$ in \nref{wavef},
we intend to show that $\{\la_n\}_{n\in\bN}$ converges to 
$\la_M$ given by (\ref{integstat0}), in $\cD'_{\mathcal{V}}(M\times M)$
in the sense of H\"ormander pseudo topology.
 According to the discussion after Definition 8.2.2 in \cite{Hormander}, this is equivalent to require that --
without assuming {\em a priori} that $\lambda_M \in \cD'_{\mathcal{V}}(M\times M)$, but assuming that
 every $\lambda_n \in \cD'_{\mathcal{V}}(M\times M)$ --
\begin{enumerate}
\item $\lambda_n\to\lambda_M$ in the topology of $\mathcal{D}'(M\times M)$,
\item $\sup\limits_n\sup\limits_{k\in V}|k|^N|\widehat{h\lambda_n}|<\infty$ for any $N\geq 1$ and for
any $h\in C^\infty_0(M\times M; \bC)$. In this last inequality $V$ stands for any cone,
closed in $\left(T^*M\right)^2\setminus 0$,
 in the 
complement of $\mathcal{V}$.
\end{enumerate}

\noindent (The former requirement is a stronger version of the result achieved in the proof of (a)
of Proposition (\ref{teo1}) where the convergence of $\lambda_n$ to $\lambda$ were proved in the 
sense of quadratic forms only.)
 The reader should notice that if  both the conditions written above were true, 
it would have to hold $\lambda_M \in \cD'_{\mathcal{V}}(M\times M)$, and this in turn implies
 $WF(\lambda_M)\subset \mathcal{V}$, a statement which {\em is nothing but 
 the inclusion $\subset$ in}  \eqref{wavef}.  Therefore the proof of the
  validity of both items above would conclude the proof of Theorem \ref{Ultima}. 
 
\noindent Let us establish the validity of both items  separately.

\proposizione\label{uniforme}{The sequence of distributions $\lambda_n\in \cD'(M\times M)$ converges to $\la_M$ in the
 weak sense.}\\

\noindent{\em Proof}.
 We have to show that, for every $h\in C^\infty_0(M\times M; \bC)$, it holds:
\beq \label{LIM}
\lim_{n\to\infty}|\la_n(h) - \la_M(h)|=0. 
\eeq
Obviously we can restrict ourselves to $h\in C_0^\infty(M\times M; \bR)$ by linearity.
In the following, we shall make use of the notations and the properties of $\Psi_h \doteq (\Ga E \otimes \Ga E) h$ 
and its Fourier transform given in the final part of
the
Appendix \ref{stimephipsi}.
The distributions $\la_n$ acts as:
\beq\label{LIM''}
\la_n(h)=\int_{\bR_+\times\bS}  \;   \left(\widehat\chi_n\otimes \widehat\chi_n  * \widehat\Psi_h\right)( -k,\om,k,
\om)\; 2k dk\; dS^2(\omega)\:,
\eeq
The Fourier transform of $F= F(\ell_1,\ell_2,\om_1,\om_2)$, 
above indicated by $\widehat{F}(k_1,k_2,\omega_1,\omega_2)$ has to be computed 
in $\bR^2$ with respect to the variable $(\ell_1,\ell_2)$ and passing to the conjugate variable $(k_1,k_2) \in \bR^2$. 
Finally it is restricted to the diagonal taking $k_1=k_2$ (and $\omega_1=\omega_2)$. Furthermore, 
$*$ stands for the convolution in $\bR^2$ so that, obviously, it results: $
\widehat{\chi_n\otimes\chi_n}  * \widehat\Psi_h=
\widehat\chi_n\otimes \widehat\chi_n  * \widehat\Psi_h = \widehat{\chi_n\otimes \chi_n \Psi_h} = 
\widehat{\chi_n \Psi_h \chi_n}$.
To justify (\ref{LIM''}), we notice that it is simply proved that the right-hand side is weakly continuous as a 
function of $h \in C^\infty_0(M\times M; \bR)$ (due to the continuity of $E\otimes E$ in the appropriate topologies 
and the presence of the 
cut-off functions  $\chi_n\otimes \chi_n$ which restrict the image of $\Gamma E\otimes \Gamma E$
to a class of functions supported in a compact subset of $\scrim\times \scrim$). On the other hand,
 when $h=f \otimes g$ with $f,g \in C_0^\infty(M;\bR)$, the right-hand side reduces to $\lambda_n(f,g)$
written as in the right-hand side of (\ref{integstat01}). Thus, by the uniqueness of the Schwartz kernel associated with 
a separately sequentially continuous quadratic form, the right-hand side of (\ref{LIM''}) individuates the distribution 
in $\cD'(M\times M)$ associated with the quadratic form $\lambda_n(\cdot,\cdot)$.
Now define the functional over $C^\infty_0(M\times M; \bR)$:
\beq\label{LMD}
\la'_M(h) \doteq \int_{\bR_+\times\bS}\; \widehat\Psi_h( -k,\om,k,\om) \; 2kdkdS^2(\omega)\:.
\eeq
{\em A priori}, there is no guarantee that it defines an element of $\cD'(M\times M)$, {\it i.e.} that it  
is weakly continuous, nor that it is the distribution associated with the quadratic form $\lambda_M$.
 However, according to (\ref{integstat0}), it results 
$\lambda'_M(f\otimes g) = \lambda_M(f,g)$. So, if $\lambda'_M$ individuated an element of $\cD'(M\times M)$,
 it would have to coincide with the distribution associated with the quadratic form $\lambda_M(\cdot,\cdot)$,
 again in view of the uniqueness of the Schwartz kernel.
 Summarising, to prove that $\lambda'_M  = \lambda_M \in \cD'(M\times M)$ it is sufficient to prove that 
\beq \label{LIM'}
\lim_{n\to\infty}|\la_n(h) - \la'_M(h)|=0. 
\eeq
In turn it would imply (\ref{LIM}).
 To prove (\ref{LIM'}) we follow a procedure similar to that employed in the second part of the 
proof  of the item (a) in Proposition \ref{teo1}. In other words 
we start considering a trivial partition of unit, constructed out of $\rho_\de$ and $\rho_\de'$ as 
follows: we choose $\rho_\de\in C^\infty_0(\bR)$ in such a way that $\rho_\de(0)=1$, $0\leq|\rho_\de(k)|\leq
1$ and $supp(\rho_\de)\subset [-\delta,\delta]$, with  $\delta >0$, whereas $\rho'_\de\doteq 1-\rho_\de$. 
Therefore
$$
\widehat\Psi_h(k_1,\om,k_2,\om')= \at \rho_\de(k_1) \rho_\de(k_2) +\rho_\de(k_1)\rho_\de'(k_2)+\rho_\de'(k_1)
\rho_\de(k_2) + \rho_\de'(k_1)\rho_\de'(k_2) \ct \widehat\Psi_h(k_1,\om,k_2,\om'),
$$
an expression we can now plug in $\lim_{n\to\infty}|\la_n(h)-\la'_M(h)|$ to get four terms: 
\begin{gather*} \int_{\bR_+ \times \bS^2}  \left|\widehat{\chi_n\Psi_h\chi_n} - \widehat{\Psi}_h\right|\;2k\; 
dk dS^2 \leq
\int_{\bR \times \bS^2}  \left|\widehat{\chi_n} * \left(\rho_\delta \widehat{\Psi}_h\rho_\delta\right)*
\widehat{\chi_n} - \rho_\delta \widehat{\Psi}_h\rho_\delta\right| \;|2k|\;dk 
dS^2 +\\
+
\int_{\bR \times \bS^2}  \left|\widehat{\chi_n} * \left(\rho_\delta \widehat{\Psi}_h\rho_\delta'\right)*
\widehat{\chi_n} - \rho_\delta \widehat{\Psi}_h\rho_\delta'\right|\;|2k|\; dk 
dS^2
+\\
\int_{\bR \times \bS^2}  \left|\widehat{\chi_n} * \left(\rho_\delta' \widehat{\Psi}_h\rho_\delta\right)*
\widehat{\chi_n} - \rho_\delta' \widehat{\Psi}_h\rho_\delta\right|\;|2k|\; dk dS^2 +
\int_{\bR \times \bS^2}  \left|\widehat{\chi_n} * \left(\rho_\delta' \widehat{\Psi}_h\rho_\delta'\right)*
\widehat{\chi_n} - \rho_\delta' \widehat{\Psi}_h\rho_\delta'\right|\;|2k|\; dk dS^2\:. 
\end{gather*}
Henceforth, we shall indicate by $A_n$,$B_n$,$C_n$ and $D_n$, respectively, the four integrals in the right-hand side.
Let us consider $D_n$. We have that, as $n\to +\infty$:
\begin{gather*}
\left(\widehat{\chi_n} * \left(\rho'_\delta \widehat{\Psi}_h\rho'_\delta 
\right)*
\widehat{\chi_n} \right)(k,p,\omega,\omega') = \\
= \int_{\bR^2}dk'dp'
\left(\widehat{\chi}(k') \left(\rho'_\delta \widehat{\Psi}_h\rho'_\delta 
\right)\at k-\frac{k'}{n},p-\frac{p'}{n},\omega,\omega'\ct
\widehat{\chi}(p') \right)
\to 
\left(\rho'_\delta \widehat{\Psi}_h\rho'_\delta 
\right)\at k,p,\omega,\omega'\ct.
\end{gather*}
Above we have used the dominated convergence
 since $\rho_\delta'\widehat{\Psi}_h\rho_\delta'= 
 \rho_\delta'\widehat{\Psi}_h\rho_\delta'(k,p,\omega,\omega')$ is in the Schwartz space 
 (the divergence has been 
 cancelled by $\rho_\delta'$) and $\widehat{\chi}$ is bounded. 
With the same argument used  in the proof second part of the 
proof  of the item (a) in Proposition \ref{teo1} to achieve (\ref{bl}), 
one sees that there is a constant $K\geq 0$ with
$$ \left|\left(\widehat{\chi_n} * \left(\rho'_\delta \widehat{\Psi}_h\rho'_\delta 
\right)*
\widehat{\chi_n} \right)(k,p,\omega,\omega') \right|\leq \frac{K}{(1+k^2)(1+p^2)}\:.$$
So, it is possible to use again Lebesgue's theorem in $D_n$,
 obtaining $D_n \to 0$  for $n\to +\infty$.\\
 Concerning $A_n$, we notice 
that it has to hold
\beq\label{AN}
|A_n| \leq\int_{\bR \times \bS^2}  
\left|\widehat{\chi_n} * \left(\rho_\delta \widehat{\Psi}_h\rho_\delta\right)*\widehat{\chi_n}\right|
\;|2k|\;dk 
dS^2
+
\int_{\bR \times \bS^2}
\left| 
\rho_\delta \widehat{\Psi}_h\rho_\delta\right| 
\;|2k|\;dk dS^2.
\eeq
The last term is bounded by
$$
 \int_{\bR\times\bR} \sp\sp dp\:dp' \int_{\bR \times \bS^2} \sp\sp dk dS^2(\omega)2|k| 
\left|\widehat{\chi_n}(-k-p)  \left(\rho_\delta \widehat{\Psi}_h\rho_\delta\right)(p, p',\om,\om)
\widehat{\chi_n}(k-p')\right|\:,
$$
Noticing that $\widehat\chi_n(p)=n\widehat\chi(np)$, and passing from the coordinates $p,p',k$ to
$u=np$, $u'=np'$, $h= nk$, the expression above can be bounded by:
$$
 \frac{1}{n^2}\int_{\bR\times\bR} \sp\sp du\:du' \int_{\bR \times \bS^2} \sp\sp dh dS^2(\omega)2(|h +u|+|u|) 
\left|\widehat{\chi}(-h-u)  \left(\rho_\delta \widehat{\Psi}_h\rho_\delta\right)\left(\frac{u}{n}, \frac{u}{n},\om,\om\right)
\widehat{\chi}(h-u')\right|\:,
$$
Using the fact that $|\widehat{\chi}|$ is bounded and has finite integral, the found integral can be bounded  by:
$$
 \frac{1}{n^2}\int_{[-n\delta,+n\delta]\times [-n\delta,+n\delta]} \sp\sp du\:du' \int_{\bS^2} \sp dS^2(\omega) 
 (K+K'|u|) \left|\widehat{\Psi}_h\left(\frac{u}{n}, \frac{u}{n},\om,\om\right)
\right|\:.
$$
 From the estimate (\ref{andamento1}), for small $k_1$ 
and $k_2$, we have $\widehat{\Psi}_h(k_1,k_2,\om,\om')\leq C/(|k_1k_2|^{Re \nu-1/2})$. This bound, inserted in the integral above
implies that, for some constants $K_1,K_2\geq 0$ independent from $\delta$:
$$
 \frac{1}{n^2}\int_{[-n\delta,+n\delta]\times [-n\delta,+n\delta]} \sp\sp du\:du' \int_{\bS^2} \sp dS^2(\omega) 
 (K+K'|u|) \left|\widehat{\Psi}_h\left(\frac{u}{n}, \frac{u}{n},\om,\om\right)
\right| \leq K_1 \delta^{3 - 2Re \nu} + K_2 \delta^{4 - 2Re \nu}\:,
$$
that is, looking back at (\ref{AN}):
$$\int_{\bR \times \bS^2}  
\left|\widehat{\chi_n} * \left(\rho_\delta \widehat{\Psi}_h\rho_\delta\right)*\widehat{\chi_n}\right|
\;|2k|\;dk 
dS^2\leq K_1 \delta^{3 - 2Re \nu} + K_2 \delta^{4 - 2Re \nu}\:.$$ 
Analogously, we find:
$$
\int_{\bR \times \bS^2} \left|\left(\rho_\delta \widehat{\Psi}_h\rho_\delta\right)\right|
\;|2k|\;dk 
dS^2(\om) \leq 
\int_{[-\delta,\delta] \times \bS^2} \left|\widehat{\Psi}_h\right|
\;|2k|\;dk dS^2(\om)
\leq 
C''\delta^{3-2Re \nu},
$$
where $C', C''\geq 0$ are constants independent on $\delta$. Therefore we have obtained that 
$|A_n|\leq H \delta^{3-2Re \nu}$, uniformly in $n$, for some constant $H\geq 0$ independent from 
$\delta$.
The remaining terms, $B_n$ and $C_n$, can be treated similarly 
making use of (\ref{andamento1}) with $n=2$,
obtaining that, 
uniformly in $n$,  $|B_n| \leq  H' \delta^{3-2Re \nu}$ 
and $|C_n| \leq  H'' \delta^{3-2Re \nu}$
for some constants $H',H''$ independent from 
$\delta$.
Following the same procedure as in the last part  of the proof of Proposition \ref{teo1}
(based on the standard properties of $\liminf$, $\limsup$), 
we can finally assert that the sequence 
of distributions
$\la_n$ tends to $\la'_M = \lambda_M\in \cD'(M\times M)$ weakly. $\Box$\\

We are now in the position to study the convergence of the 
$\la_n$ to $\la_M$ in the H\"ormander pseudo-topology $\cD'_{\mathcal{V}}(M\times M)$.
The following proposition holds,  which easily  implies the item (2) in Sec. \ref{seccentr}
when employing a partition of the unit subordinated to a covering made of domains of coordinate patches on $M\times M$
and taking  into account the compactness of the support of the $h$ appearing in the item (2) in Sec. \ref{seccentr}.

\proposizione\label{PROPLAST} {\em Let $U\subset M\times M$ a coordinate patch, $V\subset \bR^4 \times \bR^4$ 
a close conic set, so that $U\times V$ can be viewed as the corresponding portion of $T^*M\times T^*M$
employing the coordinates over $U$, and let $h\in C^\infty_0(M\times M; \bC)$ be a function supported in $U$.
If $(supp h \times V) \cap \mathcal{V} = \emptyset$, then:
\beq \label{lll}\sup\limits_n\sup\limits_{p\in V}|p|^N
|\lambda_n(e^{i\langle p,\cdot\rangle } h)|<+\infty\quad \forall  N=1,2,\ldots 
\eeq}

\noindent {\em Proof}. Obviously, by linearity, we can always assume which $h$ is real-valued and we shall
assume it henceforth. It holds $\lambda_n(e^{i\langle p,\cdot\rangle } h) = |K(e^{i\langle p,\cdot\rangle } h \otimes \chi_n \otimes \chi_n)|$
in accordance with the definition 
of the distributions $\lambda_n$ in terms of the kernel $K$ as specified in (\ref{LN}).
Thus, with $V$ as in the hypotheses,  the following inequality holds for every $N=1,2,\ldots$, because 
$p\not \in WF(\lambda_n)$ (since $WF(\lambda_n) \subset \mathcal{V}$ for (\ref{NU})):
\beq\label{maggiorazioneWF}
 |K( \chi_n \otimes \chi_n \otimes e^{i\langle p,\cdot\rangle } h)|\leq 
\frac{C_{N,n}}{(1+|p|)^N} \:,\quad \forall p \in V\:,
\eeq
for some constants $C_{N,n}\geq 0$ (depending on $h$).
 The idea is to show that, for any fixed $h\in C_0^\infty(M\times M; \bR)$ 
 there is $m_h \in \bN$, such that one can take
 $C_{N,n} = C_{N,m_h}$ constantly,    for $n\geq m_h$ in (\ref{maggiorazioneWF}).
  This fact would lead to (\ref{lll}) immediately, since it implies that:
  $$\sup\limits_n\sup\limits_{p\in V}|p|^N
|\lambda_n(e^{i\langle p,\cdot\rangle } h)| \leq \sup_n \sup_{p\in V} \frac{|p|^N C_{N,n}}{(1+|p|)^N} \leq 
\sup_n C_{N,n}\leq  \max\{C_{N,1},C_{N,2}, \ldots, C_{N,m_h}\}< +\infty\:.$$
 We need a preliminary lemma
 whose proof stays in the Appendix \ref{prooflemmafinale}. \\
 
 \lemma \label{lemmafinale} {\em If $\cO \subset \overline{\cO} \subset M$ is any open, relatively compact set, there is
 $\ell_\cO >0$ such that, viewing $\Gamma E$ a  Schwartz kernel:
 $$\mbox{sing supp}\left(\Gamma E\spa\rest_{\scrim\times \cO}\right)) \subset \cN_\cO \times \cO$$
 with
 $\cN_\cO \doteq (-\ell_\cO,\ell_\cO) \times \bS^2$ 
 in $\scrim$.} \\
 
\noindent To go on we give a precise definition of the functions $\chi_n$. As usual $\chi_n(\ell,\omega) \doteq 
\chi(\ell/n)$,
but now we define $\chi \in C_0^\infty(\bR\times \bS^2;\bR)$ as a function independent form $\omega \in \bS^2$,
 with $0\leq \chi(\ell) \leq 1$ and 
$\chi(\ell) =0$ for $|\ell|\geq 2$ whereas $\chi(\ell)=1$ for $|\ell|\leq 1$. 
Notice that the support of $\chi_n$ becomes larger and larger 
as $n$ increases and tends to cover the whole $\scrim$, taking everywhere the value $1$, as $n\to +\infty$.
Since $M$ is homeomorphic to $\bR^4\times \bR^4$, for every $h \in C_0^\infty(M\times M)$, there is a set
$\cO \doteq \cO_h$, as in the hypotheses of the Lemma \ref{lemmafinale}, such that $\cO_h\times \cO_h \supset supp\: h$.
Hence, 
there exists a sufficiently large $n_h \in \bN$ such that $\cN_{\cO_h} \subset 
supp \chi_n$ as well as $\chi_n(\overline{\cN_{\cO_h}}) = 1$
if $n\ge n_h$. \\
  We have the following bound for $K(e^{i\langle p,\cdot\rangle } h \otimes \chi_n \otimes \chi_n)$:
\begin{gather}
|K(\chi_n \otimes \chi_n \otimes e^{i\langle p,\cdot\rangle }h )|\leq
\nonumber
\\
|K(\chi_{n_h} \otimes \chi_{n_h} \otimes e^{i\langle p,\cdot\rangle } h)|+
|K((\chi_{n}-\chi_{n_h}) \otimes \chi_{n_h}\otimes e^{i\langle p,\cdot\rangle } h)|+
\nonumber
\\
|K(\chi_{h_n} \otimes (\chi_{n}-\chi_{n_h})\otimes e^{i\langle p,\cdot\rangle } h)|+
|K((\chi_{n}-\chi_{n_h}) \otimes (\chi_{n}-\chi_{n_h})\otimes e^{i\langle p,\cdot\rangle } h)|
\label{terminiWF}
\end{gather}
Let us indicate by $P_n$, $Q_n$ and $R_n$ the first the third and the last term, respectively, in the right-hand side of the inequality above. The second can be discussed similarly to the third. 
Let us analyze the features of $P_n$, $Q_n$ and $R_n$ separately proving that there exist a natural  $m_h$ which guarantees the validity of the thesis as discussed above, for each term separately, thus
concluding the proof.

{\bf Analysis of $P_n$}. The analysis  is straightforward because $n=n_h$ is fixed.
As $p\in V$ does not belong to the $WF(\la_{n_h})$, for every $N$ it exist $C_{N,n_h}\geq 0$ such that 
$$
P_n = |K(\chi_{n_h} \otimes \chi_{n_h} \otimes e^{i\langle p,\cdot\rangle } h)| \leq \frac{C_{N,n_h}}{(1+|p|)^N},\qquad \forall p\in V,\quad \forall N\geq 1.
$$ 

{\bf Analysis of $R_n$}.   With our definition 
of $n_h$, the function $\chi_{n}-\chi_{n_h}$ vanishes over $\overline{\cN_\cO}$ for $n\geq n_h$. 
Thus, 
due to the Lemma \ref{lemmafinale}, the wave front set of 
$((\chi_{n}-\chi_{n_h})\Gamma E \otimes (\chi_{n}-\chi_{n_h})\Gamma E)\spa\rest_{\scrim\times \scrim \times \cO\times \cO}$ is  empty, 
so that every $p$ in $\bR^4\times\bR^4$ individuates a direction of rapid decrease for 
it. As $supp h\subset \cO_h$, this result allows us to estimate  the rate of rapid decreasing of 
$K((\chi_{n}-\chi_{n_h}) \otimes (\chi_{n}-\chi_{n_h})\otimes e^{i\langle p,\cdot\rangle } h)$.
In the Appendix \ref{prooflemmafinale}, we shall prove the following lemma

\lemma \label{lemmafinale2}
 {\em With the hypotheses of Proposition \ref{PROPLAST}, for every $N=1,2,\ldots,$ there is a constant $C_N\geq 0$ such that:
\begin{gather}
|k_1|^2 \; 
|k_2|^2 \; 
|p|^N
\left|(\Ga E \otimes \Ga E) ((\chi_{n}-\chi_{n_h})e^{i\langle k_1,\cdot\rangle}  \otimes (\chi_{n}-\chi_{n_h})e^{i\langle k_2,\cdot\rangle} \otimes  h e^{i\langle p,\cdot\rangle})
\right| \leq C_N \label{addedW}\:,
\end{gather}
 when $n \geq n_h$.} \\

\noindent Let us prove that (\ref{addedW}) implies that, if $n\geq  n_h$:
\beq \label{addedW2}
R_n = |K((\chi_{n}-\chi_{n_h}) \otimes (\chi_{n}-\chi_{n_h}) \otimes e^{i\langle p,\cdot\rangle})|
\leq  \frac{C_N}{(1+|p|)^N},\qquad \forall p\in V\:, \forall N \geq 1\:,
\eeq
for some $C_N \geq 0$. Indeed, (\ref{addedW}) entails that, for every $N$, it exist a $C'_N\geq 0$, which does not depend on $n \geq n_h$, such that:
\begin{gather}
\left|(\Ga E \otimes \Ga E) ((\chi_{n}-\chi_{n_h})e^{i\langle k_1,\cdot\rangle}  \otimes (\chi_{n}-\chi_{n_h})e^{i\langle k_2,\cdot\rangle} \otimes h e^{i\langle p,\cdot\rangle})
\right|
\leq 
\frac{1}{(1+|k_1|)^2(1+|k_2|)^2} \frac{C'_N}{(1+|p|)^N}. \label{ST}
\end{gather}
The left-hand side is nothing but $H^2\left|\left(\widehat{(\chi_{n}-\chi_{n_h})} \otimes \widehat{(\chi_{n}-\chi_{n_h})}
*\widehat{\Psi}_{h e^{i\langle p,\cdot\rangle}}\right)(k_1,\omega,k_2,\omega')\right|$.
Therefore, from the very definition of the kernel $K$,
 $K((\chi_{n}-\chi_{n_h}) \otimes (\chi_{n}-\chi_{n_h}) \otimes e^{i\langle p,\cdot\rangle})$ is obtained 
by integrating $(\Ga E \otimes \Ga E) ((\chi_{n}-\chi_{n_h})e^{i\langle k_1,\cdot\rangle}  \otimes (\chi_{n}-\chi_{n_h})e^{i\langle k_2,\cdot\rangle} \otimes h e^{i\langle p,\cdot\rangle})$  with $k_1=-k_2 =k$ and $\omega =\omega'$ 
 over $\bR_+\times \bS^2$ with respect to the measure $kdk dS^2(\omega)$.
 In that way, (\ref{ST}) yields that $C_N\geq 0$ exists  such that, $\forall n \geq n_h$:
$$
|K((\chi_{n}-\chi_{n_h}) \otimes (\chi_{n}-\chi_{n_h}) \otimes e^{i\langle p,\cdot\rangle})|
\leq 4\pi \left(\int_{0}^{+\infty} \sp\sp \frac{2k dk}{(1+|k|)^4}\right)  \frac{C'_N}{(1+|p|)^N} =  \frac{C_N}{(1+|p|)^N},\qquad \forall p\in V\:, \forall N \geq 1\:.
$$
We have finally proved the validity of (\ref{addedW2}).

{\bf Analysis of $Q_n$}. We start by:
$$
Q_n=|K(\chi_{n_h} \otimes (\chi_{n}-\chi_{n_h})\otimes e^{i\langle p,\cdot\rangle } h))| \leq
|K(\chi_{n_h} \otimes (\chi_{n}-\chi_{3n_h})\otimes e^{i\langle p,\cdot\rangle } h)| 
+
|K(\chi_{n_h} \otimes (\chi_{3n_h}-\chi_{n_h})\otimes e^{i\langle p,\cdot\rangle } h)|.
$$
First of all notice that the wave front set of
$WF(K(\chi_{n_h} \otimes (\chi_{3n_h}-\chi_{n_h})\otimes \cdot  ))\subset \mathcal{V}$,
 as $K(\chi_{n_h} \otimes (\chi_{3n_h}-\chi_{n_h})\otimes \cdot  )$ can be seen as a 
 composition of distribution with compact support and the Theorem 8.2.14 of 
 \cite{Hormander} can be applied twice. Therefore, every $p\in V$ is a direction of rapid
 decreasing for such distribution and the rate of decrease does not depend on $n$ by construction.
Let us pass to analyse the first term in the right-hand side of the inequality written above for $Q_n$. Notice that 
 the support of $\chi_{n_h}$ never intersects the support of $(\chi_n- \chi_{3n_h})$ if $n\geq 3n_h$, hence the 
 singularity $(\ell-\ell')^{-2}$ present inside $K$ due to $T$ (see \nref{distributionT} and Proposition \ref{PT})
  is harmless. For this reason we can skip the $\epsilon$-prescription present in $T$ (reminded by $0^+$ in \nref{distributionT}) and we can consider the part of the integral kernel of $T$ depending on $\ell$ and $\ell'$ as a smooth function.
We pass  to establish the existence of a $n$-uniform bound for 
$
|p|^N|K(\chi_{n_h} \otimes  (\chi_{n}-\chi_{3n_h})\otimes e^{i\langle p,\cdot\rangle } h)|
$. We have the bound, where, as before, $y=(y_1,y_2)$ and with obvious notation concerning derivatives:
\begin{gather}
|p|^N
|K( \chi_{n_h} \otimes  (\chi_{n}-\chi_{3n_h})\otimes e^{i\langle p,\cdot\rangle } h )|
\leq
\nonumber
\\
\leq
\left|
\int\sp  d\ell dS^2(\omega) d\mu(y_1)
 \Ga E(\ell,\om,y_1)\int \sp d\mu(y_2) e^{i\langle p, y\rangle} D_{y}^{N} h(y) \chi_{n_h} (\ell) \int 
 \sp \mathcal{E}(\ell',\om,y_2)
 \frac{\chi\at\frac{\ell'}{n}\ct
-\chi_{3n_h} (\ell')}{(\ell-\ell')^2} 
  d\ell' \right|.
 \label{ultimastima}
\end{gather}
Notice that due to the domain property of $h$ and $\chi(\ell'/n)
-\chi_{3n_h} (\ell')$, the associated causal propagator is a smooth function
similarly to the discussion done for the term $R_n$ mande in the proof of Lemma \ref{lemmafinale2}. 
As we done in the proof of that lemma, we have denoted it by $\mathcal{E} (\ell',\om',y_2)$
 in the formula
above, where the two internal integrations have the standard meaning, whereas the external one has to be understood 
in the distributional sense. To find an estimate for the right-hand side of (\ref{ultimastima})
 it is convenient to define, for $n= 3n_h, 3n_h +1,\ldots, \infty$:
$$
F^{\beta}_n(\ell,\om,y) \doteq \rho(y)\int_{\bR} D_{y_2}^{\beta} \mathcal{E} (\ell',\om,y_2) \chi_{n_h}(\ell) \frac{\aq\chi\at\frac{\ell'}{n}\ct
-\chi_{3n_h} (\ell') 
\cq}{(\ell-\ell')^2} d\ell'$$
and 
$$
F^{\beta}_\infty (\ell,\om,y) \doteq \rho(y)\int_{\bR} D_{y_2}^{\beta} \mathcal{E} (\ell',\om,y_2)\chi_{n_h}(\ell) \frac{
1-\chi_{3n_h} (\ell')}{(\ell-\ell')^2} d\ell'\:, \quad G_n^\beta(y)\doteq  \rho(y)^t(\Ga E)( F^\beta_n)(y)\:.
$$
where $\rho(y) = \rho_1(y_1)\rho_2(y_2)$ is such that $\rho_1,\rho_2 \in C_0^\infty(M;\bR)$ and $\rho(y)=1$ if $y \in \overline{\cO_h \times \cO_h}$.
Notice that, because of the decay property at large $\ell'$ of $\mathcal{E}$ discussed 
in the analysis of $R_n$,  and due of the boundedness of $\chi(\cdot/n)
-\chi_{3n_h} (\cdot)$, all the $F^\beta_n$ as well as $F^\beta_\infty$ are smooth functions. So they belong to 
$C^\infty_0(\scrim \times M \times M; \bR)$ by construction. By direct inspection one proves that
$F^{\beta}_n \to F^{\beta}_\infty$ in the topology of $C^\infty_0(\scrim \times M \times M; \bC)$, as $n\to +\infty$.
Furthermore, in view of Theorem 8.2.12 and the discussion before Theorem 8.2.13 in \cite{Hormander} (and we adopt here the notation used therein), as $WF(^t(\Ga E))_M$ is the empty set, it results that $G_n$ 
is smooth  on $M\times M$ for every $n \geq 3n_h$ including $n=\infty$, it is compactly supported within a $n$-independent relatively compact set including $\overline{\cO_h}$,
in view of the presence of
$\rho$, and $G_n \to G_\infty$ in the topology of
$C_0^\infty(M\times M)$ as $3n_h \leq n\to +\infty$. Equipped with the introduced functions, 
 the right hand side of (\ref{ultimastima}) can be rearranged as
$$\left| \int 
 e^{i \langle p, y \rangle}  D^{N}_{y} G^0_n(y) h(y)  \:
d \mu(y_1) d\mu(y_2)
\right|\:,$$ 
so that 
we finally have:
\begin{gather*}
|p|^N
|K( \chi_{n_h} \otimes  (\chi_{n}-\chi_{3n_h})\otimes e^{i\langle p,\cdot\rangle } h )|
\leq
\sum_{\beta+\beta_1+\beta_2=N}
\int 
\left|D^{\beta_2}_{y_2} \at G^\beta_n(y)  D^{\beta_1}_{y_1} h(y)\ct  
\right| 
d \mu(y_1) d\mu(y_2)
\leq C'_{N,n}
\end{gather*}
However, since $G_n \to G_\infty$ in the topology of
$C_0^\infty(M\times M)$ as $3n_h \leq n\to +\infty$, there exists $C_N < +\infty$ with $C'_{N,n} \leq C_N$, so that,
 for $n\geq 3n_h$:
$$
Q_n=|K( \chi_{n_h} \otimes (\chi_{n}-\chi_{n_h}) \otimes e^{i\langle p,\cdot\rangle } h)| \leq \frac{C_N}{(1+|p|)^N},\qquad \forall p\in V,\quad \forall N\geq 1
$$
Finally, collecting the estimates for the four terms in \nref{terminiWF} we get that \nref{maggiorazioneWF} 
holds true with $C_{N,n} = C_{N,m_h}$ constantly if $n\geq m_h$ whenever one assumes
 $m_h\doteq 3n_h$, and this conclude the proof. $\Box$.\\

\noindent As discussed in Sec. \ref{seccentr}, the achieved result implies the inclusion $\subset$
in  \eqref{wavef} and this concludes the proof of Theorem \ref{Ultima}. 

\se{Conclusion}
In \cite{DMP} we established that, out of a bulk-to-boundary reconstruction procedure, it is possible to 
identify a preferred quasifree algebraic state for a scalar field theory living on any manifold lying in a 
large class of Friedmann-Robertson-Walker spacetimes.
The first goal accomplished in the present paper is the extension of those results to other physically relevant cases, encompassing the analyses  of the linear scalar fluctuations of the metric in inflationary models.
As a further result, in this paper, we proved that such state is also 
of Hadamard form. 
This entails several interesting consequences the most notable being both the boundedness 
of the back-reaction due to quantum effects and the possibility to perform over such a state a 
renormalisation procedure. Despite the interest of these remarks, we should emphasise once more that the 
possibly most interesting application of our results lies in inflationary models according to which the early 
Universe undergoes an almost de Sitter phase of expansion driven by a real scalar field coupled to a 
self-interaction potential. Within this framework many consequences are derived employing quantisation and 
perturbative techniques and the most notable is the existence of a power spectrum of fluctuations which is 
almost scale invariant. A byproduct of this paper is indeed the possibility to put on a firmer mathematical 
ground the often taken for granted assertion that it exists a well-behaved quantum state out of which the 
above results can be derived.

Nonetheless we should point out that our proof holds true only under same further hypothesis and most notably
$\nu$, taken as in \eqref{nu}, must differ from $\frac{3}{2}$ which would correspond to a massless scalar
field minimally coupled to scalar curvature. In this scenario the techniques employed in this paper cannot be
applied and we reckon that a rethinking of the whole procedure is necessary. Nonetheless we feel safe to 
claim that the proof of the Hadamard condition, as it stands now, is of physical interest from the point of 
view of cosmology and it opens the road to tackle specific inflationary models on a firm mathematical ground.

\section*{Acknowledgements.} The work of C. D. was supported by the von Humboldt Foundation and, that of N. P., 
by the German DFG Research Program SFB 676. 
 Part of this paper was written by V. M. and N. P. at the E. Schr\"odinger Institute of Vienna during the ESI Program 
{\em Operator Algebras and Conformal Field Theory}. V. M. and N. P. are therefore grateful to the institute and the organisers of the program
Y. Kawahigashi, R. Longo, K.-H. Rehren and J. Yngvason for their kind hospitality.
The authors would like to thank K. Fredenhagen and R. Brunetti for 
useful discussions.\\

\appendix

\section{\label{approxsol}Approximated solution by means of perturbation theory}

We shall briefly comment both on the extension of the construction of the solution $\chi_k(\tau)$ of \nref{rhokeq} 
in the case $\nu \in \bR$
with
$1/2<\nu<3/2$ and $V(\tau)=O(\tau^{-5})$ as well as on the related uniform estimates that we have used throughout. The values 
$0\leq \nu\leq 1/2$ (or $\nu$ imaginary) were considered in \cite{DMP2}.
The construction is similar to the one presented in the proof of theorem 4.5 in 
\cite{DMP2} though some subtleties arise. To be more specific, in the mentioned theorem, $\chi_k$ was 
constructed by means of a perturbative series around a particular solution $\chi_k^0(\tau)$ which corresponds 
to \eqref{mode}, that in the de Sitter background. More precisely
\begin{gather}
\chi_k(\tau) = \chi^0_k(\tau)\notag\\
+  \spa\sum_{n=1}^{+\infty}(-1)^{n}\spa \int_{-\infty}^{\tau}\sp\sp\sp dt_1
\int_{-\infty}^{t_1}\sp\sp\sp dt_2 \cdots  \int_{-\infty}^{t_{n-1}}\sp\sp\sp \sp dt_n
S_k(\tau,t_1)S_k(t_1,t_2)\cdots S_k(t_{n-1},t_n) V(t_1)V(t_2)\cdots V(t_n)\chi^0_k(t_n),\label{seriepert}
\end{gather}
where 
\beq
S_k(t,t'):= -i \left(\overline{\chi^0_k(t)}\chi^0_k(t')- \overline{\chi^0_k(t')}\chi^0_k(t)\right)\:,\quad 
t,t'\in (-\infty,0)\:, \label{Sk}
\eeq
is the retarded fundamental solution of the unperturbed one dimensional problem \nref{rhokeq}, whereas $V(
\tau)$ is the perturbation potential. In the case under investigation, the proof of the convergence of the 
series \nref{seriepert} can be proved, dividing the problem into two parts, namely we shall discuss the cases 
$0\leq k<1$ and $k\geq 1$ separately. Though the former is the most difficult, we can nonetheless take into
account the following behaviour for $|k\tau|\leq1$:
$$
|\chi^0_k(\tau)| 
\leq \frac{C_\nu}{2}  
\sqrt{\tau} \at |k\tau|^{\nu} + |k\tau|^{2-\nu}+\frac{1}{|k\tau|^\nu}\ct
, \qquad |\tau| > 1,\; |k\tau|<1
$$
as well as that for $|k\tau|\geq 1$ 
$$
|\chi^0_k(\tau)| 
\leq \frac{C_\nu}{\sqrt{|k|}} 
, \qquad |\tau| > 1,\; |k\tau|\geq 1\;.
$$
Together they imply the following $\tau$-uniform estimate
\beq\label{stimachi}
|\chi^0_k(\tau)| 
\leq 
\frac{C_\nu}{2}  
\at \frac{1}{|k|^{\nu}} +\frac{1}{\sqrt{|k|}}\ct
, \qquad |\tau| > 1\;
\eeq
where $C_\nu$ is some positive $\nu$-dependant constant. The analysis of $S_k(t_1,t_2)$ is more subtle, we 
shall derive two estimates valid for small and large $k$ respectively. More precisely
$$
|S_k(t_1,t_2)| \leq  C_\nu' \at \sqrt{|t_1t_2|} \ct (|t_1|^\nu|t_2|^{2-\nu}+|t_1|^\nu|t_2|^
{2-\nu}), \leq C'_\nu{|t_1|}^{2\alpha +1}, 
\;\; |t_1| > |t_2|>1,  |k|<1\; 
$$
where $0<\alpha\doteq\sup\{\nu,2-\nu\}<\frac{3}{2}$, while
$$
|S_k(t_1,t_2)| \leq  C_\nu'  \frac{1}{|k|},
\qquad |t_1| > |t_2|>1,\qquad |k|\geq 1\; 
$$
 $C_\nu'$ being a further constant different from $C_\nu$. Thanks to these two estimates and since the 
perturbation potential is $V(\tau)=O(\tau^{-5})$, the series \nref{seriepert} is dominated by the following 
convergent series 
$$
|\chi_k(\tau)|\leq  |\chi^0_k(\tau)|\sum_{n=0}^\infty \frac{1}{n!} \at \frac{C}{(3-2
\alpha)\: |\tau|^{3-2\alpha}} \ct^n,\; |k|<1 \; 
\quad 
|\chi_k(\tau)|\leq
|\chi^0_k(\tau)|
\sum_{n=0}^\infty \frac{1}{n!} \at \frac{C}{4|k| |\tau|^{4}} \ct^n,   
\; |k|\geq 1.  
$$
with a fixed constant $C$. The two inequalities show that the behaviour in $k$ of $|\chi_k(\tau)|$ is similar 
to the one presented in \eqref{stimachi} for $|\chi^0_k(\tau)|$.
 It is possible to prove that the $\tau$ derivatives do not alter the bound given at small $k$. Notice that similar estimates hold true also 
for $\pa_\tau\chi_k(\tau)$. More precisely, since for $|k|<1$, both $S_k(t_1,t_2)$ and $\chi_0(\tau)$ can be expanded in powers of $k$ the following bound must hold for the $n$th order $\tau$ derivative of $\chi_k(\tau)$ 
\beq\label{stimapartialchi}
\partial_{\tau}^n\chi_k(\tau)
=
\frac{C^{n+1}_\nu}{2}  
 \frac{1}{|k|^{\nu}} + O({ |k|^{\min(\nu,2-\nu)}} )
, \qquad |\tau| > 1\;, |k|<1 \;.
\eeq
These estimates are valid for $\nu \in \bR$ with $0\leq \nu < 3/2$. For $\nu \in i\bR$ there is no singularity at $k=0$.
To conclude this appendix we would like to emphasise that, as $\tau\to -\infty$, 
the analogy between $\chi_k^0$ and $\chi_k$ becomes stronger; most notably, if we rearrange the series 
\eqref{seriepert} and we operate as before, in the limit $\tau \to\infty$, we end up with
$$
|\chi_k(\tau)-\chi_k^0(\tau)|\leq C_\nu \at \frac{1}{|k|^{\nu}} +\frac{1}{\sqrt{|k|}}\ct
\at {e^{C/\tau^{3-2\alpha}}-1}\ct  \to 0, \qquad \tau \to -\infty\;.
$$

\section{\label{stimephipsi}Derivation of some estimates both for $\Phi_h$ and $\Psi_h$}

In this appendix we derive some estimates both for the bi-solutions of Klein-Gordon equation 
 generated by real smooth compactly-supported 
smooth on $M\times M$ and for their restrictions to the horizon $\scrim\times\scrim$.\\
To this end, take $h\in C^\infty_0(M\times M; \bR)$ and define the  two-wavefunction $\Phi_h$ and its extension/restriction 
$\Psi_h$ to $\scrim\times \scrim$ (referring to the larger globally-hyperbolic spacetime $(\widehat{M}, \widehat{g}_{FRW})$
including  $(M,g_{FRW})$ as a subspace):
$$
\Phi_h \doteq \left(E\otimes E\right)\; h\:, \quad \Psi_h \doteq H^{-2}\left[\left(\widehat{E}\otimes \widehat{E}\right)\; h\right]\rest_{\scrim\times \scrim}\:.
$$
It result $\Phi_h\in 
C^\infty(M\times M; \bR)$ and similarly $\Psi_h\in 
C^\infty(\scrim \times \scrim; \bR)$. It follows
by direct application of Theorems 8.2.9 and 8.2.12 in \cite{Hormander}
and taking into account the shape of $WF(E)$ (\ref{WFE}). 
Moreover, it turns out that the restriction of $\Phi_h$ to $\Sigma_\tau \times\Sigma_\tau$ has compact support when
$\Sigma_\tau$ is a constant-time Cauchy surface of $(M,g_{FRW})$. To prove it, take an open set $A\subset M$
such that $\overline{A}$ is compact and $supp h \subset \overline{A}\times \overline{A}$ (such $A$ does exist because
$M$ is homeomorphic to $\bR^4$). Thus consider class of functions $f_n,g_n \in C_0^\infty(A;\bR)$ with
$\sum_{n=1}^N f_n\otimes g_n \to h$ in $C_0^\infty(A\times A; \bC)$ as $N \to +\infty$.
By the known properties of $E$ in globally hyperbolic spacetime, the restriction of $Ef$ 
to a spacelike Cauchy surface $\Sigma$ is included in 
the compact $J(H) \cap \Sigma$ when $f \in C^\infty(M; \bR)$ is such that 
$supp f \subset H$ and $H$ is compact.
(Here and henceforth $J(A)\doteq J^+(A)\cup J^-(A)$.) As a consequence, 
the restriction of $(E\otimes E)\left( \sum_{n=1}^N f_n\otimes g_n\right)
= \sum_{n=1}^N E(f_n)\otimes E(g_n)$ to $\Sigma_\tau\times \Sigma_\tau$
has support included in the compact $B\doteq (J(\overline{A}) \cap \Sigma_\tau) \times (J(\overline{A}) \cap \Sigma_\tau)$.
Since  $\sum_{n=1}^{N} E(f_n)\otimes E(g_n) \to h$  uniformly on every compact set, as $N\to +\infty$, and each
$(E(f_n)\otimes E(g_n))\spa \rest_{\Sigma_\tau\times \Sigma_\tau}$ vanishes outside $B$, we have that for every
 compact subset $B'$ of
$\Sigma_\tau\times \Sigma_\tau$ with  $B' \supset B$, it also holds
$h\rest_{B'\setminus B} =0$. Taking  $B'$ larger and larger (this is possible because $\Sigma_\tau$ is homeomorphic
to $\bR^3$) one finds that $\Phi_h\spa \rest_{\Sigma_\tau\times \Sigma_\tau}=0$ outside the compact $B$, so that 
$supp\left( \Phi_h\rest_{\Sigma_\tau\times \Sigma_\tau} \right)= supp \left[\left( (E\otimes E)h\right)\rest_{\Sigma_\tau\times \Sigma_\tau}\right] \subset B$ is compact.\\
 $\Phi_h$ can be decomposed into modes 
along the lines of \eqref{general}, formally:
$$
\Phi_h(x_1,x_2)=
\int\limits_{\bR^3\times \bR^3}
\left[
\phi_{\bf k_1}(x_1)
\phi_{\bf k_2}(x_2)
\widetilde\Phi_h(\bk_1,\bk_2)+
\overline{
\phi_{\bf k_1}(x_1)
\phi_{\bf k_2}(x_2)
\widetilde\Phi_h(\bk_1,\bk_2)}\; \right] d^3\bk_1 d^3\bk_2.
$$ 
Let us analyse the properties of $\widetilde\Phi_h(\bk_1,\bk_2)$ and its dependence on $h$. To 
this end, fix any, but fixed, initial time $\tau_0$ and the associated Cauchy surface $\Si_{\tau_0}
$. Since $\Phi_h$ is smooth by construction, we can restrict it on $\Si_{\tau_0}\times\Si_{\tau_0}$.
Hence the previous formula is invertible by means of a procedure similar to that of \nref{modes}:
\begin{gather}
\widetilde\Phi(\bk_1,\bk_2)=-i\int\limits_{\Si_{\tau_0}\times \Si_{\tau_0}}\: d^3x_1 d^3x_2\; a^4(\tau_0)
\left[
\frac{\partial\overline{\phi_{\bk_1}(x_1)}}{\partial\tau_1}
\frac{\partial\overline{\phi_{\bk_2}(x_2)}}{\partial\tau_2}
\Phi_h(x_1,x_2)+
\right.\notag
\\ 
\label{bimodes}\left.-
\frac{\partial\overline{\phi_{\bk_2}(x_2)}}{\partial\tau_2}
\overline{\phi_{\bk_1}(x_1)}\frac{\pa\Phi_h(x_1,x_2)}{\partial
\tau_1}
\overline{\phi_{\bk_1}(x_1)}
\overline{\phi_{\bk_2}(x_2)}
\frac{\pa^2\Phi_h(x_1,x_2)}{\pa \tau_1\partial
\tau_2}
-\frac{\partial\overline{\phi_{\bk_1}(x_1)}}{\partial\tau_1}
\overline{\phi_{\bk_2}(x_2)}\frac{\pa\Phi_h(x_1,x_2)}{\partial
\tau_2}
\right].
\end{gather}
In combination both with the explicit form of the modes $\phi_\bk$ as in \nref{general} and with $\Phi_h,
\partial_{\tau_1}\Phi_h,\partial_{\tau_2}\Phi_h\in C_0^\infty\left(\Si_{\tau_0}\times\Si_{\tau_0}; \bC\right)$, such 
expression entails that $\widetilde\Phi_h(\bk_1,\bk_2)$ is an integrable function which is smooth
except for $\bk_1=0$ and $\bk_2=0$ separately. Furthermore it decays 
rapidly at large $\bk_1$ or $\bk_2$ uniformly in the angles (when the other variable is fixed) whereas, near $\bk_1=0$ and $\bk_2=0$ 
it has the following angle-independent uniform 
bound:
$$
|\widetilde\Phi_h(\bk_1,\bk_2)| \leq \frac{C}{(|\bk_1||\bk_2|)^{Re\nu}}.
$$
The behaviour can be summarised as it follows $\widetilde\Phi_h=\widetilde\Phi_h(\bk_1,\bk_2)$
is everywhere smooth but $\bk_1=0$ and $\bk_2=0$ separately, moreover,
 for $n=1,2, \ldots$ there are constants $C_n\geq 0$ with:
\beq\label{andamento}
|\widetilde\Phi_h(\bk_1,\bk_2)| \leq \frac{C_n}{(|\bk_1||\bk_2|)^{Re\nu}} \at\frac{1}{(1+|\bk_1|+|\bk_2|)^n}\ct, 
\quad \mbox{for all $\bk_1\bk_2 \in \bR^3\setminus\{0\}$,}
\eeq
finally, the constants $C_n$ depend continuously on $h$ with respect to the topology of compactly supported smooth functions 
on $M\times M$. This last observation can be proved using the continuity of the Fourier transform on 
$\Si_{\tau_0}\times\Si_{\tau_0}$ with respect to the Schwartz topology
 and the continuity of the causal propagator in the appropriate topologies, remembering 
that the restriction to $\Si_{\tau_0}\times\Si_{\tau_0}$ of the wavefunction $\Phi_h =(E\otimes E)h$ is compactly supported.\\
We can now pass to consider the smooth restriction of $\Phi_h$ to $\scrim\times \scrim$, 
$\Psi_h$. Adapting a procedure similar to that we
exploited for the wavefunctions, it arises that:
$$
\Psi_h(\ell,\om,\ell',\om') = H^5 i \int_{\bR \times \bR} \frac{e^{-i\ell k_1-i\ell' k_2}}{2\pi}  \sqrt{\frac{k_1 k_2 }{4}} 
\widetilde\Phi_h(H k_1,\eta(\om),H k_2,\eta(\om')) dk_1 dk_2.
$$
\noindent The bound \nref{andamento} entails some properties of 
$
\widehat\Psi_h( k_1,\om,k_2,\om')\doteq  iH^5
\sqrt{\frac{k_1 k_2 }{4}} 
\widetilde\Phi_h(H k_1,\eta(\om),H k_2,\eta(\om')).
$
The function 
$\widehat\Psi_h=\widehat\Psi_h( k_1,\om,k_2,\om')$ is everywhere smooth, except for $k_1=0$ or/and $k_2=0$, moreover
the following angle-uniform bound holds, for every $n=1,2,\ldots,$
\beq\label{andamento1}
|\widehat\Psi_h( k_1,\om,k_2,\om')| \leq \frac{C_n'}{|k_1 k_2|^{Re \nu-1/2}} \at\frac{1}{(1+|k_1|+|k_2|)^n}\ct, 
\eeq
where each constant $C_n'$ depends continuously on $h$ as before. Accordingly, it holds that 
$\widehat\Psi_h( k_1,\om,k_2,\om')$ is 
an integrable function on $\bR\times\bS^2\times \bR \times \bS^2$.

\section{\label{prooflemmafinale} Proofs of some propositions}

\noindent{\bf Proof of Lemma \ref{lemmafinale}}. 
As $E$ is the restriction to $M$ of the analogous causal propagator $\widehat{E}$ defined in the 
larger spacetime $\widehat M$ including $M$ as subspacetime,
and since the singular support of $\widehat{E}\in \cD'(\widehat{M} \times \widehat M)$ is given by the points 
$(y,x)\in \widehat{M} \times \widehat M$ such that there is a null $\widehat{g}$-geodesic connecting them, the lemma
is proved if we establish that the null $\widehat{g}$-geodesic getting out from the compact 
$K \doteq \overline{\cO}\subset M$ intersect $\scrim$ in a compact set.\\
Let us prove this fact. 
To this end, as in \cite{DMP2} and referring to the spacetime  $(M,g_{FRW})$ with coordinates and metric 
as in  (\ref{FRWconf1}),
 we introduce the new null coordinates $U=\tan^{-1}(\tau+r)$ and 
$V=\tan^{-1}(\tau-r)$ ranging in subsets of $\bR$ individuated by $\tau \in (-\infty,0)$ and $r\in (0,+
\infty)$.  Then:
\beq\label{metcomp}
g_{FRW}=\frac{a^2(\tau(U,V))}{\cos^2 U\cos^2 V}\left[-\frac{1}{2}dU\otimes dV - \frac{1}{2}dV \otimes dU+
\frac{\sin^2(U-V)}{4}d\bS^2(\theta,\varphi)\right].
\eeq
$M \cup \scrim$ is the wedge $|V| \geq |U|$ with $V \in [-\pi/2,0]$, but
removing the boundary at $U=-V$ (including its endpoints) corresponding to $\tau= +\infty$,  and omitting the point 
$(-\pi/2,-\pi/2)$, corresponding to the tip of $\scrim$ which does not exist in $\widehat{M}$. In this picture $\scrim$ coincides with the boundary 
at $V= -\pi/2$ (without its endpoints), whereas 
the apparent boundary at $U=V$ is the submanifold  $r=0$. However, with our choice of $a$, the metric 
in (\ref{metcomp}) is smooth outside this region, too.
As a matter of fact, the globally hyperbolic spacetime $(\widehat M, \widehat g)$, 
which extends $(M,g_{FRW})$ is obtained  letting $V$ ranging in a neighbourhood of $\scrim$ including a region beyond it.
 The relevant
 point here is that also the metric $\widetilde{g}$ obtained cancelling the overall factor, 
$a^2(\tau(U,V))/(\cos^2 U\cos^2 V)$, in the right-hand side of 
 (\ref{metcomp}), is well-behaved and  smooth for  $U,V \in \bR$ with $U\geq V$ (the apparent singularity for $U=V$ 
is a coordinate singularity only).  The spacetime $(\widetilde{M}, \widetilde{g})$ obtained in that way is nothing but
the (globally hyperbolic) Einstein static universe. The remarkable point is that, within this picture,
 $\scrim$ coincides with $\partial J^+(i^-; \widetilde{M})\setminus\{i^-\}
 =\partial I^+(i^-; \widetilde{M})\setminus\{i^-\} $, where the point  $i^- \in \widetilde{M}$ 
is the tip of the cone $\scrim$ localised at $U=-\pi/2$, $V= -\pi/2$. 
Furthermore $M = I^+(i^-; \widetilde{M})$.
An immediate consequence is that no future-directed  null (or causal)  $\widetilde{g}$-geodesics emanating from $K$ can reach $\scrim$
since $J^+(K;\widetilde{M}) \subset I^+(i^-; \widetilde{M})$ which is always open and so, it cannot intersect $\partial I^+(i^-; \widetilde{M}) = \scrim \cup \{i^-\} $.
Since the two metrics are conformally related \cite{Wald},
the null geodesics of $(\widetilde{M},\widetilde{g})$
when restricted to $(\widehat M \cup \scrim,\widehat g)$ individuate $\widehat g$-geodesics and 
{\em vice versa}. Therefore
our thesis
would proved if we were able to establish that the {\em past-directed} null $\widehat{g}$-geodesic getting out from the compact 
$K \doteq \overline{\cO}\subset M$ intersect $\scrim$ in a compact set (notice that $\scrim$ does not includes $i^-$).\\
Actually that result is true, since the analogous statement, for the opposite time orientation, 
 was established in \cite{M2} in the proof of Lemma 4.3 
for a generic globally
hyperbolic spacetime $\widetilde{M}$, when $M\subset \widetilde{M}$ is the globally hyperbolic 
subspacetime  $I^-(i^+; \widetilde{M})$ with $i^+ \in \widetilde{M}$ and $\scri\doteq \partial I^-(i^+; \widetilde{M})
 \setminus \{i^+\} = \partial J^-(i^+; \widetilde{M}) \setminus\{i^+\}$ 
 has the same geometric structure as $\scrim$  (referred 
 to the opposite temporal orientation). In that case $K\subset M$ was any compact set (the further hypotheses 
  assumed for $K$  in the proof of Lemma 4.3 in \cite{M2} played no role in the part of the proof we are interested in).  
 $\Box$\\

\noindent{\bf Proof of Lemma \ref{lemmafinale2}}. In the following,
 $\mathcal{E}(\ell,\om,y)$ is the smooth integral kernel of 
$\Ga E$ with the left entry restricted to $\scrim\setminus \cN_{\cO_h}$ and the right one restricted to $\cO_h$.  
 To achieve (\ref{addedW}), we start from:
\begin{gather}
|k_1|^2 \; 
|k_2|^2 \; 
|p|^N
\left|(\Ga E \otimes \Ga E) ((\chi_{n}-\chi_{n_h})e^{i\langle k_1,\cdot\rangle}  \otimes (\chi_{n}-\chi_{n_h})e^{i\langle k_2,\cdot\rangle} \otimes  h e^{i\langle p,\cdot\rangle})
\right|
\leq
\nonumber
\\ 
\;
\sp \sp
\sum_{\alpha+\beta+\beta'=N}
\int_{M \times M}\sp \sp \sp
d\mu(y,y')   
|D_y^\alpha h(y)|
\int_{\bS^2} d\om
\left| 
\int_\bR d\ell\; e^{ik_1 \ell}
\pa_\ell^2
D_y^\beta\mathcal{E}(\ell,\om,y_1)
\; \at\chi\at\frac{\ell}{n}\ct-\chi_{n_h} (\ell)\ct\; 
\right| \cdot
\nonumber
\\
\cdot
\left|
\int_\bR d\ell'
e^{ik_2 \ell'} 
\pa_{\ell'}^2
D_y^{\beta'}\mathcal{E}(\ell' ,\om,y_2)
\; \at\chi\at\frac{\ell'}{n}\ct-\chi_{n_h} (\ell' )\ct\;   
\right|
.
\label{impazzisco}
\end{gather}
where $y=(y_1,y_2)\in \cO_h\times \cO_h$,
$d\mu(\ell,\ell',\omega,\omega',y,y') \doteq  d\ell dS^2(\omega)d\ell'dS^2(\omega')d\mu_g(y)d\mu_g(y')$ and, finally,
 $D_y$ are shortcut notations for the derivatives along the coordinates of $y$. 
To prove  (\ref{addedW}), since the domains of integration in $y,y'$ and $\omega$ in the right-hand side of (\ref{impazzisco}) have finite measure, it is enough proving that each of the two internal integrals in $d\ell$ and $d\ell'$ give rise to
functions of the remaining variables which are uniformly bounded in $n$. In other words we have now to establish that:
\beq\label{fine5}
 \left| 
\int_\bR d\ell\; e^{ik_1 \ell}
\partial_\ell^2 
D_y^\beta\mathcal{E}(\ell ,\om,y_1)
\; \at \chi\at\frac{\ell}{n}\ct-\chi_{n_h} (\ell )\ct\;   
\right|\leq C_\beta<+\infty\:, \quad \mbox{uniformly in $n=1,2,\ldots$.}
\eeq
  To prove (\ref{fine5}) we study the behaviour
 of the smooth function $\pa_{\ell}^n
D_y^{\beta}\mathcal{E}(\ell,\om,y)$. Starting form the analysis of the solutions of the Klein-Gordon equation 
in terms of modes summarised in Sec.\ref{secmodes} and using results in the subsequent Sec.\ref{secproj}, it  arises that the causal propagator can be written, in the sense of the distributional $\epsilon$-prescription, i.e.  smearing the kernel with a test function $g=g(y)$ before taking the limit,
 as
\beq \label{Propsmooth}
D_y^\beta \partial^m_\ell\mathcal{E}(\ell,\om,y) = \lim_{\epsilon\to 0^+} c\: Im
 \left\{D_y^\beta \partial^m_\ell\int_0^{+\infty} dk e^{i3\pi/4}e^{i k  \ell/H } \sqrt{k}   \chi_k(\tau)  e^{-ik |\vec{y}| \cos(\omega,\vec{y})} 
e^{-\epsilon|k|}\right\}\:,
\eeq
where $c\in \bR$ is a constant irrelevant in our discussion, $y \doteq (\tau,\vec{y})$  and $\cos(\omega,\vec{y})$ 
is a shortcut notation for the cosine of the angle between  $\vec{y}$ and the unit vector individuates by the
 angles $\omega \in \bS^2$ when adopting spherical coordinate on the Cauchy surface $\Si_\tau$ the 
 functions $\chi_k$ are the modes discussed in Sec.\ref{secmodes}.  We have written $\mathcal{E}$ instead of $\Gamma E$,
 because we are intersted in the case $(\scrim \setminus \cN_{\cO}) \times \cO_h \ni ((\ell, \omega), y)$ which 
 implies that the kernel of $\Gamma E$ is smooth.
We decompose the integral above into two parts $\int_0^{+\infty} = \int_0^1 + \int_1^{+\infty}$ and we define
$$D_y^\beta \partial^m_\ell\mathcal{E}_<(\ell,\om,y) \doteq \lim_{\epsilon\to 0^+} c\: Im
 \left\{D_y^\beta \partial^m_\ell\int_0^{1} dk e^{i3\pi/4}e^{i k  \ell/H } \sqrt{k}   \chi_k(\tau)  e^{-ik |\vec{y}| \cos(\omega,\vec{y})} 
e^{-\epsilon|k|}\right\}\:,$$
and
$$D_y^\beta \partial^m_\ell\mathcal{E}_>(\ell,\om,y) \doteq \lim_{\epsilon\to 0^+} c\: Im
 \left\{D_y^\beta \partial^m_\ell\int_1^{+\infty} dk e^{i3\pi/4}e^{i k  \ell/H } \sqrt{k}   \chi_k(\tau)  e^{-ik |\vec{y}| \cos(\omega,\vec{y})} 
e^{-\epsilon|k|}\right\}\:.$$

{\bf The case of $\mathcal{E}_<(\ell,\om,y)$}.
The limit as $\epsilon\to 0^+$ for first integral can be computed without using a smearing test function and the limit 
can be intechanged with the symbol of integral. This arises by direct application of Lebesgue's dominated convergence theorem.
Furthermore, defining $a:= (\tau + \ell/ H - k |\vec{y}| \cos(\omega,\vec{y}) )$, we have the bound, following from
 (\ref{stimapartialchi}) when $m=\beta=0$:
\begin{align}
&\lim_{\epsilon \to 0^+}\left|c \:Im \int_0^1 dk e^{i3\pi/4}e^{i k  \ell/H } \sqrt{k}  
 \chi_k(\tau)  e^{-ik |\vec{y}| \cos(\omega,\vec{y})} e^{-\epsilon|k|}\right|
 = \left|c\int_0^1 dk e^{i3\pi/4}e^{i k  \ell/H } \sqrt{k}  
 \chi_k(\tau)  e^{-ik |\vec{y}| \cos(\omega,\vec{y})}\right| \nonumber\\
 &\leq
 |c|a^{Re \nu -3/2} \frac{C_\nu}{2} \int_0^a u^{1/2-Re \nu} du
+ |c|a^{-3/2 - \min(Re \nu, 2)} \int_0^a O(k^{1/2 + \min(Re \nu, 2)}) du\:, \label{aggW1}
\end{align}
where, in our hypotheses $|a|>0$ since $|\ell|$ is very large wheras $y$ and $\omega$ range in a bounded domain
(in other words $((\ell,\om), y)$ does not belong to the singular support of $\Gamma E$).
The function in the second line of (\ref{aggW1}) vanishes as $|\ell| \to +\infty$ uniformly in 
$y \in \cO_h$ and $\omega \in \bS^2$.
An analogous procedure can easily be implemented in presence ot derivatives $D_y^\beta, \partial^m_\ell$, making use 
of (\ref{stimapartialchi})
again. The final result is that, for both $\nu \in i\bR$ or $\nu \in (0,3/2)$, each function:
$$(\scrim \setminus \cN_{\cO}) \times \cO_h \ni ((\ell, \omega), y) \mapsto D_y^\beta \partial^m_\ell\mathcal{E}_<(\ell,\om,y)$$
 vanishes as $|\ell| \to +\infty$ uniformly in $(\omega, y) \in \bS^2 \times \cO_h$, so that it is bounded. Furthermore, if $m>0$,
 it is also $\ell$ integrable and the integral is bounded as a function of $(\omega,y) \in \bS^2 \times \cO_h$. \\
Looking at the left-hand side of (\ref{fine5}), we have the bound concerning the only contribution due to 
$D_y^\beta\mathcal{E}_<(\ell ,\om,y)$:
\begin{gather}
\left| 
\int_\bR d\ell\; e^{ik \ell}
\partial_\ell^2 
D_y^\beta\mathcal{E}_<(\ell ,\om,y)
\; \at \chi\at\frac{\ell}{n}\ct-\chi_{n_h} (\ell )\ct\;   
\right| \nonumber 
\\
\leq 
\int_\bR d\ell\; \left|
\partial_\ell^2 
D_y^\beta\mathcal{E}_<(\ell ,\om,y)
\right|
\; 
\left| \chi\at\frac{\ell}{n}\ct-\chi_{n_h} (\ell )  
\right| 
+
2
\int_\bR d\ell\;
\left| 
\partial_\ell
D_y^\beta\mathcal{E}_<(\ell ,\om,y)
\right|
\; 
\left| \partial_\ell\chi\at\frac{\ell}{n}\ct-\partial_\ell\chi_{n_h} (\ell )  
\right| 
+\nonumber 
\\
+
\int_\bR d\ell\;
\left| 
D_y^\beta\mathcal{E}_<(\ell ,\om,y)
\right|
\; 
\left| \partial_\ell^2 
\chi\at\frac{\ell}{n}\ct-\partial_\ell^2 
\chi_{n_h} (\ell )  
\right| \:.\label{addedW3}
\end{gather}
Let us start by analyzing the third integral in the right-hand side. Performing the change of variables $\ell \to n
\ell$, it can be rewritten as:
\beq \int_\bR d\ell\;
\left| 
D_y^\beta\mathcal{E}_<(\ell ,\om,y_1)
\right|
\; 
\left| \partial_\ell^2 
\chi\at\frac{\ell}{n}\ct-\partial_\ell^2 
\chi_{n_h} (\ell )  
\right| =
\int_\bR d\ell\;
\left| D_y^\beta\mathcal{E}_<(n\ell ,\om,y_1)\right|\; 
n^{-1}\left| \partial_\ell^2 \chi\at\ell \ct-\partial_\ell^2 \chi_{n_h} (n\ell )  \right|\:.\label{addedW23}
\eeq
and the right-hand side is $n$-uniformly bounded 
by the product of $\sup_{(y,\omega)\in \cO_h\times \scrim} |D_y^\beta\mathcal{E}_<|$ --
which we know to be finite -- and
$$\frac{1}{n}\int_\bR d\ell\;
\left| \partial_\ell^2 
\chi\at\ell\ct-\partial_\ell^2 
\chi_{n_h} (n\ell )  
\right| \leq \frac{1}{n}
\int_\bR d\ell\;
\left| \partial_\ell^2 
\chi (\ell )  
\right| +
\int_\bR ds\;
\left|
\partial_s^2 
\chi_{n_h}\at s\ct
\right|  \leq  D < +\infty\:, \quad \forall n=1,2,\ldots\:.
$$
The remaining two terms in the right hand side of 
(\ref{addedW3}) can be $n$-uniformly bounded similarly. The second term can be treated 
with an analogous procedure, with the change of variables $\ell \to n\ell$
using the fact that $\partial_\ell
D_y^\beta\mathcal{E}_<(\ell ,\om,y)$ is bounded on $(\scrim \setminus \cN_{\cO_h})\times \cO_h$ and that
it holds 
$n^{-1}\int_\bR d\ell\; 
\left| \partial_\ell\chi\at\ell\ct-\partial_\ell\chi_{n_h} (n\ell ) \right| <G<+\infty$, 
uniformly in $n$. The first one can be treated analogously, noticing that
 $\left| \chi\at\frac{\ell}{n}\ct-\chi_{n_h} (\ell )  
\right| $ is bounded uniformly $n$, whereas 
$\int_{\bR \setminus [-\ell_{\cO_h}, \ell_{\cO_h}]} d\ell\;
\left| 
\partial_\ell^2 
D_y^\beta\mathcal{E}_< (\ell ,\om,y)
\right| \leq H <+\infty $ if $y\in \cO_h$.\\
We have established that:
$$
 \left| 
\int_\bR d\ell\; e^{ik_1 \ell}
\partial_\ell^2 
D_y^\beta\mathcal{E}_<(\ell ,\om,y_1)
\; \at \chi\at\frac{\ell}{n}\ct-\chi_{n_h} (\ell )\ct\;   
\right|\leq C <+\infty\:, \quad \mbox{uniformly in $n=1,2,\ldots$}
$$
To conclude it is enough to establish the analog for $\mathcal{E}_>$.

{\bf The case of $\mathcal{E}_>(\ell,\om,y)$}. We pass now to study the behaviour of 
$\mathcal{E}_>(\ell,\om,y)$.
As before we first examine the case $m=\beta=0$.
To this end we exploit an approximation procedure to compute the modes $\chi_k(\tau)$ for $k>1$ which is
 similar to that used in \cite{DMP2} and discussed in the Appendix \ref{approxsol}, but now using a different decomposition of 
 the complete potential $k^2 + a(\tau)^2\left[m^2 + \left(\xi- \frac{1}{6} \right)R(\tau) \right]$ into ground  and perturbation 
 parts, $W_0(k)$ and $W(\tau)$, respectively. In fact, we
  define $W_0(k)\doteq k^2$ and $W(\tau)\doteq a(\tau)^2\left[m^2 + \left(\xi- \frac{1}{6} \right)R(\tau) \right]$
   in the differential equation in (\ref{rhokeq}), so that the equation  now reads:
\begin{gather}
 \frac{d^2}{d\tau^2}\chi_k(\tau)  + (W_0(k)  + W(\tau)) \chi_k(\tau) =0\:. \label{rhokeq2}
\end{gather}   
We stress that the modes $\chi_k$ are the same as that found employing the differential equation (\ref{rhokeq}), only the perturbative procedure to solve it
is different. The solution can be written as in (\ref{seriepert}) where now, 
$\chi_k^0(\tau)\doteq\frac{e^{-i\pi/4}}{\sqrt{2 k}} e^{-ik\tau}$, 
$V$ is replced by $W$, and $S_k(t,t')$ is replaced by:
$
 T_k(t,t')\doteq\frac{\sin(k(t-t'))}{2k}
$.
By direct inspection one sees that the found series can be re-arranged as:
\beq\label{serie2}
e^{i\pi 4}\sqrt{k}\chi_k(\tau) = \sum_{n=0}^\infty \at\frac{-1}{ik}\ct^n \aq A^+_n(k,\tau) e^{ik\tau} +  
A^-_n(k,\tau) e^{-ik\tau} \cq\:, 
\eeq
where the coefficient $A^\pm_n$ satisfy:
$A^+_0(k,\tau)\doteq 0$, $A^-_0(k,\tau)\doteq 1$
and  recursive relations:
\begin{gather*}
A^+_{n+1}(k,\tau)\spa = \int_{-\infty}^\tau\spa \sp\sp W(t)\aq A^+_n(k,t)+ \spa A^-_n(k,t) e^{-2ik t}  \cq  dt\;,
\quad
A^-_{n+1}(k,\tau)\spa = -\int_{-\infty}^\tau \spa\sp\sp W(t)\aq A^-_n(k,t)+ \spa A^+_n(k,t) e^{2ik t}  \cq  dt. 
\end{gather*}
All the integrand involved in the recursive procedure are absolutely integrable, due to the form of $W$ and
the series (\ref{serie2}) turns out to be uniformly absolutely convergent with its $t$-derivatives for $(t,k) \in I(t_0)\times (1,+\infty)$, where $I(t_0)$ is a neighborhood of every $t_0<-T$, 
so that it can be derived under the symbol of series and, in this way, one can check that the left-hand side
individuates a solution of (\ref{rhokeq2}). 
All the coefficient $A^\pm_n(k,\tau)$ are bounded uniformly in $k$. To be more precise, 
the $k$ dependence of those coefficient appears as  oscillating phases under some sign of integration. 
Furthermore there is always a part of $A^-_n(k,t)$ that does not depend on $k$ at all,
 while the coefficient $A^+_n(k,t)$ explicitly depends on $k$.
We are now in place to discuss the behaviour of $\mathcal{E}_>(\ell,\om,y)$ (always remaining  out of the singularities
of  $\mathcal{E}(\ell,\om,y)$,
i.e. for $y \in \cO_h$ and $(\ell,\omega)\in \scrim\setminus
\cN_{\cO_h}$).
We have, for $y=(\tau,\vec{y})$,
\begin{gather*} \nonumber
\pa_\ell^m D^\beta_y \mathcal{E}_>(\ell,\om,y) 
=
\lim_{\epsilon\to 0^+} c\:  Im  \left\{
\pa_\ell^m D^\beta_y\spa 
\int_1^\infty \sp\sp\spa dk \;i\; e^{i k  \ell/H } 
 \sum_{n=0}^\infty \at\frac{-1}{ik}\ct^n \aq A^+_n(k,\tau) e^{ik\tau'} +  A^-_n(k,\tau) e^{-ik\tau'} \cq\spa
e^{-\epsilon|k|}\right\}\:,
\end{gather*}
where 
$\tau'\doteq \tau - |\vec{y}| \cos(\omega,\vec{y}) $.
As before, we start by considering the case $m=0$ and $\beta=0$. 
The terms of the expansion of $\mathcal{E}_>$ that could give rise to problems 
at large $k$ are only those with $n=0$ and $n=1$, the remaining rest $O(1/k^2)$ produces $(y,\omega)$-uniformly bounded functions after evaluation of the integral (it can be evaluated without the $\epsilon$-prescription ). So, let us to examine:
\begin{gather} 
\lim_{\epsilon\to 0^+} c\: Im  \left\{
\pa_\ell^m D^\beta_y
\int_1^\infty \sp\sp dk \;i\; e^{i k  \ell/H } 
 \aq e^{-ik \tau'} +
\at\frac{-1}{ik}\ct  \at  A^+_1(k,\tau) e^{ik\tau'} +  A^-_1(k,\tau) e^{-ik\tau'} \ct\cq
e^{-\epsilon|k|}\right\}\:,\label{decap}
\end{gather}
where $m=\beta=0$ and
\begin{gather*}
A^+_{1}(\tau,k):= \int_{-\infty}^\tau W(t) \; e^{-2ik t}  \;  dt\;,
\qquad
A^-_{1}(\tau,k):= -\int_{-\infty}^\tau W(t)\;  dt, 
\end{gather*}
and thus $A^-_1$ does not depend on $k$, while 
$A^+_1$ has an oscillating phase inside an integral.
The first term in the integral (\ref{decap}) gives rise to a distribution proportional to:
\beq\label{distdelta}
 \de{(\ell/H-\tau + |\vec{y}| \cos(\omega,\vec{y}))} - \frac{\sqrt{2}}{\sqrt{\pi}} 
 \frac{\sin(\ell/H-\tau + |\vec{y}| \cos(\omega,\vec{y}))}{(\ell/H-\tau + |\vec{y}| \cos(\omega,\vec{y}))} \:.
\eeq
In this formula, actually, the Dirac delta cannot contribute because it is supported outside the domain we are considering,
moreover the remaining term is  bounded uniformly in $(y,\omega) \in \cO_h\times \bS^2$
and it falls off as  $1/\ell$ at large $\ell$.
Let us now consider the term in the integral arising from $A^-_1$. It looks like
$$
A(\tau) \int_1^\infty \frac{\sin(k(\ell/H-\tau))}{k} dk 
$$
which decays, for large $|\ell|$,
faster then $1/|\ell|^{\alpha}$, with $0\leq\alpha<1$ so, it is $(y,\omega)$-uniformly bounded.
The term containing $A^+_1$ can also be easily shown to be $(y,\omega)$-bounded.
That term can be written as 
$$
\lim_{\epsilon\to0^+} c\; Im \ag\int_1^{\infty} dk\; i\; e^{-\epsilon |k| }\int_{-\infty}^\tau e^{ik( \tau'+\ell/H-2t)} W(t) \cg.
$$
Performing the $k$-integration, a distribution like \nref{distdelta} apperas, under the sign of integration in $t$ and it can also be shown to be bounded.  
Furthermore considering $m_1$ $y$-derivatives and $m_2$ $\ell$-derivatives of $\mathcal{E}_>$ does not affect its uniform boundedness, in particular, the extra terms of that series $n<m_1+m_2+2$ that needs to be taken into account, they can always treated as one of the case studied before, using the fact that the derivatives of 
the distribution \nref{distdelta} are either supported outside the 
support of $\mathcal{E}_>$ or vanish uniformly for large $\ell$.
 
We conclude that, for $m=0,1$ and any $\beta$, every function:
$$(\scrim \setminus \cN_{\cO}) \times \cO_h \ni ((\ell, \omega), y) \mapsto D_y^\beta \partial^m_\ell\mathcal{E}_>(\ell,\om,y)$$
 vanishes as $|\ell| \to +\infty$ uniformly in $(\omega, y) \in \bS^2 \times \cO_h$, so that it is bounded.
Looking at the left-hand side of (\ref{fine5}), we have the bound concerning the only contribution due to 
$D_y^\beta\mathcal{E}_>(\ell ,\om,y)$:
\begin{gather}
\left| 
\int_\bR d\ell\; e^{ik \ell}
\partial_\ell^2 
D_y^\beta\mathcal{E}_>(\ell ,\om,y)
\; \at \chi\at\frac{\ell}{n}\ct-\chi_{n_h} (\ell )\ct\;   
\right| \nonumber 
\leq
\\
\leq 
\left|\int_\bR d\ell\; e^{ik \ell} \left(\partial_\ell^2 D_y^\beta\mathcal{E}_>(\ell ,\om,y) \right) 
 \at \chi\at\frac{\ell}{n}\ct-\chi_{n_h} (\ell )  \ct \right| 
+
2
\int_\bR d\ell
\left| 
\partial_\ell
D_y^\beta\mathcal{E}_>(\ell ,\om,y)
\right|
\left| \partial_\ell\chi\at\frac{\ell}{n}\ct-\partial_\ell\chi_{n_h} (\ell )  
\right| 
+\nonumber 
\\
+
\int_\bR d\ell\;
\left| 
D_y^\beta\mathcal{E}_>(\ell ,\om,y)
\right|
\; 
\left| \partial_\ell^2 
\chi\at\frac{\ell}{n}\ct-\partial_\ell^2 
\chi_{n_h} (\ell )  
\right| \:.\label{addedW3F}
\end{gather}
As before, we start by analyzing the third integral in the right-hand side. Performing the change of variables $\ell \to n
\ell$, it can be rewritten as:
\beq \int_\bR d\ell\;
\left| 
D_y^\beta\mathcal{E}_>(\ell ,\om,y_1)
\right|
\; 
\left| \partial_\ell^2 
\chi\at\frac{\ell}{n}\ct-\partial_\ell^2 
\chi_{n_h} (\ell )  
\right| =
\int_\bR d\ell\;
\left| D_y^\beta\mathcal{E}_<(n\ell ,\om,y_1)\right|\; 
n^{-1}\left| \partial_\ell^2 \chi\at\ell \ct-\partial_\ell^2 \chi_{n_h} (n\ell )  \right|\:.\label{addedW23F}
\eeq
and the right-hand side is $n$-uniformly bounded 
by the product of $\sup_{(\omega,y)\in \bS^2\times\cO_h} |D_y^\beta\mathcal{E}_>|$ --
which we know to be finite -- and
$$\frac{1}{n}\int_\bR d\ell\;
\left| \partial_\ell^2 
\chi\at\ell\ct-\partial_\ell^2 
\chi_{n_h} (n\ell )  
\right| \leq \frac{1}{n}
\int_\bR d\ell\;
\left| \partial_\ell^2 
\chi (\ell )  
\right| +
\int_\bR ds\;
\left|
\partial_s^2 
\chi_{n_h}\at s\ct
\right|  \leq  D < +\infty\:, \quad \forall n=1,2,\ldots\:.
$$
 The second term in (\ref{addedW3F}) can be treated 
with an analogous procedure, with the change of variables $\ell \to n\ell$
using the fact that $\partial_\ell
D_y^\beta\mathcal{E}_>(\ell ,\om,y)$ is bounded on $(\scrim \setminus \cN_{\cO_h})\times \cO_h$ and that it also holds 
$n^{-1}\int_\bR d\ell\; 
\left| \partial_\ell\chi\at\ell\ct-\partial_\ell\chi_{n_h} (n\ell ) \right| <G<+\infty$, 
uniformly in $n$.\\
To conclude, in order to establish that:
$$
 \left| 
\int_\bR d\ell\; e^{ik \ell}
\partial_\ell^2 
D_y^\beta\mathcal{E}_>(\ell ,\om,y)
\; \at \chi\at\frac{\ell}{n}\ct-\chi_{n_h} (\ell )\ct\;   
\right|\leq C <+\infty\:, \quad \mbox{uniformly in $n=1,2,\ldots$}
$$
so concluding the overall proof, it is sufficient to prove a $n$-uniform bound 
 for the first term in the right-hand side of (\ref{addedW3F}):
$$
\left|\int_\bR\; e^{ik\ell} \at \pa_\ell^2\;  D_y^\beta\; \mathcal{E}_>(\ell,\om,y) \ct
\at\chi\at\frac{\ell}{n}\ct-\chi_{n_h}\ct
d\ell 
\right| \leq C(\beta)\:, \quad\mbox{uniformly in $n$ and in $k$.}
$$
Without loosing generality we shall substitute the $y$ derivative with $ik$ factors in the $k$-expansion of $\mathcal{E}_>$. The terms not considered here are harmless, in particular, the case when the $y$ derivative is a $\tau$ derivative that acts on the first of the recursive integrals in the perturbative series can only lower the degree of divergence.
%
We use the expansion found beforehand with $w=\beta+2$:
\begin{gather*}
\lim_{\epsilon\to 0^+}
\left|\int_\bR\; e^{ik\ell}
 Im  \left\{ 
\int_1^\infty dk \;i\; e^{i k  \ell/H } (ik)^w 
 \sum_{m=0}^\infty \at\frac{-1}{ik}\ct^n \aq A^+_m(k,\tau') e^{ik\tau'} +  A^-_m(k,\tau') e^{-ik\tau'} \cq
e^{-\epsilon|k|}\right\}\:
\right.
\cdot
\\
\cdot
\left.
\at\chi\at\frac{\ell}{n}\ct-\chi_{n_h}\ct
d\ell 
\right|
\end{gather*}
The coefficient $A_m^-(k,\tau')$ decompose into the sum 
$A_m^-(k,\tau)= B_m(\tau') + B^-_m(k,\tau')$,
where $B_m(\tau')$ does not depend on $k$.
We start considering only these $k$-independent terms.
\begin{gather*}
\lim_{\epsilon\to 0^+}
\left|\int_\bR\; e^{ik\ell}
\left\{
\int_\bR   dk \; \Theta(|k|-1)\;(ik)^w  e^{i k  \ell/H } 
 \sum_{m=0}^\infty \at\frac{-1}{ik}\ct^n B_m(\tau) e^{-ik\tau'} 
e^{-\epsilon|k|}\right\}\:
\right.
\cdot
\\
\cdot
\left.
\at\chi\at\frac{\ell}{n}\ct-\chi_{n_h}\ct
d\ell 
\right|.
\end{gather*}
Notice that the preceding integral can be seen as a
the convolution between $\widehat{\mathcal{E}}$ and $\widehat\chi_n-\widehat\chi_{n_h}$,
\begin{gather*}
\lim_{\epsilon\to 0^+}
\left|
\int_\bR  \; \Theta(|q|-1)\;  
 \sum_{m=0}^\infty (-1)^m (iq)^{w-m} B_m(\tau) e^{-iq\tau} 
e^{-\epsilon|q|}\:
\at n \widehat\chi\at n(k-q)\ct-\widehat\chi_{n_h}(k-q)\ct
dq 
\right|.
\end{gather*}
We now divide the  sum above as $\sum_{w-m\geq 0} + \sum_{w-m< 0}$ and analyze the second sum. 
\begin{gather*}
\lim_{\epsilon\to 0^+}
\left|
\int_\bR  \; \Theta(|q|-1)\;  
 \sum_{m=w+1}^\infty (-1)^m (iq)^{w-m} B_m(\tau) e^{-iq\tau} 
e^{-\epsilon|q|}\:
\at n \widehat\chi\at n(k-q)\ct-\widehat\chi_{n_h}(k-q)\ct
dq
\right| \leq \\
\leq
 \sum_{m=w+1}^\infty | B_m(\tau') |
\int_\bR  \;    
\left|
\:
 n \widehat\chi\at n(k-q)\ct\right| + \left| \widehat\chi_{n_h}(k-q)
\right|
dq 
 = C(w)
\end{gather*}
where, in the last term, we have used the fact that the series of the continuous functions $\tau'\mapsto |B_n(\tau')|$ converges uniformly so that the sum is continuous and thus it admits a finite bound when
$(\omega, y) \in \bS^2\times \cO_h$ (remind that $\tau' = \tau -|\vec{y}\cos(\omega,y)|$),
 and that the last integral does not depend both on $q$ and $n$.
The first term in the series can be treated analogously exploiting the fact that 
$q^m\Theta(|q|-1)=q^m-q^m\Theta(1-|q|)$
and noticing that $q^m$ is the Fourier transform of the $m$th-derivative of the delta distribution supported outside the support of $\chi_n-\chi_{n_h}$, hence it cannot give any contribution.
Finally, it remains  to consider:
\begin{gather*}
\lim_{\epsilon\to 0^+}
\left|
\int_\bR  \; \Theta(1-|q|)\;  
  \sum_{m=0}^{w} (-1)^m (iq)^{w-m} B_m(\tau') e^{-iq\tau} 
e^{-\epsilon|q|}\:
\at n \widehat\chi\at n(k-q)\ct-\widehat\chi_{n_h}(k-q)\ct
dq 
\right| \leq \\
\leq
 \sum_{m=0}^{w} | B_m(\tau') |
\int_\bR  \;    
\left|
\:
 n \widehat\chi\at n(k-q)\ct\right| + \left| \widehat\chi_{n_h}(k-q)
\right|
dq 
 = C'(w) 
\end{gather*}
The remaining term present in the perturbative series, those associated with $B^-_n(q,\tau)$ and $A^+_n(q,\tau)$, can be treated analogously, apart for the contribution of the delta functions.
It is  in fact not always true that the support of the derivatives of the deltas stay  outside the support of the $\chi$s, but in this case, the delta distributions appear inside one of the recursive integrals, similarly to the case of the proof of boundedness of $A^+_1(k,t)$ considered before, hence they are harmless.

\end{document}